\documentclass[10pt,oneside,reqno]{amsart}
\usepackage{amsfonts,amssymb, amscd,amsmath,latexsym,amsbsy,bm}
\numberwithin{equation}{section}
\usepackage{mathrsfs}
\usepackage{dsfont}
\usepackage{braket}
 \usepackage[usenames,dvipsnames]{xcolor}
 \usepackage[titletoc]{appendix}
\usepackage{multirow}

\usepackage[colorlinks,hyperindex, 
           bookmarks=true,bookmarksopen=true]{hyperref}
           \hypersetup    {
        colorlinks=true,
       linkcolor=black,
        urlcolor=MidnightBlue,
        citecolor=RedOrange,
   }

\usepackage{epic}
\usepackage{eepic}
\usepackage{graphicx}

\usepackage{pgf}
\usepackage{tikz}

\usetikzlibrary{shapes}
\usetikzlibrary{plotmarks}

\usepackage{cite}
\usepackage{enumerate}
\usepackage[curve]{xy}
\usepackage{stmaryrd}

\usepackage{mathtools}

\renewcommand{\author}[1]{\large\rm #1\\ \bigskip}
\renewcommand{\title}[1]{\bigskip\bigskip\Large\bf #1\bigskip\bigskip\\}

\newcommand{\Bigpsi}[3]{\phantom{\Psi}_2 \kern -.05em
\Psi_2\left(\genfrac{}{}{0pt}{}{#1}{#2}\biggl|#3\right)}

\newcommand{\bea}{\begin{eqnarray}}
\newcommand{\eea}{\end{eqnarray}}

\newcommand{\beq}{\begin{equation}}
\newcommand{\eeq}{\end{equation}}

\renewcommand{\textcolor}[1]{}

 {\begin{list}{}%
         {\setlength{\leftmargin}{#1}}%
         \item[]%
 }
 {\end{list}}

\usepackage{mathtext}
\usepackage{stackrel}

\usepackage[utf8]{inputenc}

\begin{document}

$\;$\\
\vspace{2.7cm}

\begin{center}
{\LARGE \bf From well-known tensor mesons to yet unknown axial-tensor mesons }
\end{center}

\vspace{0.2cm}

\vskip 1.3 cm

\centerline{\large {\bf Shahriyar Jafarzade$^{a}$\footnote{
\href{mailto:shahriyar.jzade@gmail.com}{shahriyar.jzade@gmail.com}}, Arthur Vereijken$^{a}$\footnote{
 \href{mailto:Arthur.vereijken@gmail.com}{Arthur.vereijken@gmail.com}}, Milena Piotrowska$^{a}$\footnote{
 \href{mailto:milena.piotrowska@ujk.edu.pl}{milena.piotrowska@ujk.edu.pl}} and Francesco Giacosa$^{a,b}$\footnote{
 \href{mailto:fgiacosa@ujk.edu.pl}{fgiacosa@ujk.edu.pl}}}  }

{\small
\begin{center}
\textit{ $^{a}$Institute of Physics, Jan Kochanowski University,\\ ul. Uniwersytecka 7, 25-406,, Kielce, Poland} , \\
\vspace{2.5mm} 
\textit{ $^{b}$ Institute for Theoretical Physics, J. W. Goethe University,\\
Max-von-Laue-Str. 1, 60438 Frankfurt am Main, Germany} \\
\texttt{} \\
\vspace{.1mm}
\vspace{.1mm}
\end{center}
}

\vskip 0.5cm \centerline{\bf Abstract} \vskip 0.2cm \noindent 

While the ground-state tensor ($J^{PC}=2^{++}$) mesons $a_{2}(1320)$, $K_{2}^{\ast}(1430)$, $f_{2}(1270)$, and $f_{2}^{\prime}(1525)$ are well known experimentally and form an almost ideal nonet of quark-antiquark states, their chiral partners, the ground-states axial-tensor ($J^{PC}=2^{--}$) mesons are poorly settled: only the kaonic member $K_2(1820)$ of the nonet has been experimentally found, whereas the isovector state $\rho_2$ and two isoscalar states $\omega_2$ and $\phi_2$ are still missing.  
Here, we  study masses, strong, and radiative decays of tensor and axial-tensor mesons within a chiral model that links them: the established tensor mesons are used to test the model and to determine its parameters, and subsequently various predictions for their chiral partners, the axial-tensor mesons, are obtained.  The results are compared to current lattice QCD outcomes as well as to other theoretical approaches and show that the ground-state axial-tensor mesons are expected to be quite broad, the vector-pseudoscalar mode being the most prominent decay mode followed by the tensor-pseudoscalar one. 
Nonetheless, their experimental finding seems to be possible in ongoing and/or future experiments.

\newpage

{\color{black}
\small
\vspace{-0.45cm}
{\parskip=0pt
\tableofcontents
}
\vspace{-0.25cm}
}
\normalsize

\section{Introduction}

\label{sec:introduction}

The classification of the observed mesonic resonances into nonets of quark-antiquark ($q\bar{q}$) states using flavor symmetry (as well as chiral symmetry) is one  of the greatest achievements of modern high energy physics. 

In the non-relativistic spectroscopic notation, $q\bar{q}$ nonets are specified by the expression $n ^{2S+1}L_J$, with $n$ being the radial quantum number, and $S$, $L$ and $J$ being the spin, spacial, and total angular momenta of the quark-antiquark system, respectively. Upon restricting to the first radial excitation $n=1$, the situation is summarized in Table \ref{mesonlist}, in which also the relativistic notation $J^{PC}$ (with parity $P=(-1)^{L+1}$ and charge conjugation $C=(-1)^{L+S}$) is reported.

Examples of very well-known $q\bar{q}$ mesonic nonets include the pseudoscalar mesons with $J^{PC}=0^{-+}$ (spectroscopic notation: $^{1}S_{0}$),
the vector mesons 
with $J^{PC}=1^{--}$ ($^{3}S_{1}$), 
and the axial-vector mesons
with $J^{PC}=1^{++}$ ($^{3}P_{1}$) \cite{Zyla:2020zbs}.  
The above resonances are in good agreement with the quark model of Ref. \cite{isgur1985} and the Bethe-Salpeter approach of Ref. \cite{Fischer:2014xha}.
The scalar mesons 
with $J^{PC}=0^{++}$ ($^{1}S_{0}$) are in the centre of a long-standing debate, e.g. Ref. \cite{Pelaez:2015qba,Sarantsev:2021ein,Rodas:2021tyb,Binosi:2022ydc,Klempt:2022qjf,Klempt:2021nuf,Klempt:2021wpg,Guo:2022xqu}: there is no lack of candidates, on the contrary there are more states than expected from $q\bar{q}$ alone, in agreement with the intrusion of four-quark states and the scalar glueball.
Summarizing, upon including the nonet of pseudovector mesons with $J^{PC}=1^{+-}$ ($^{1}P_{1}$) and the orbitally excited vector mesons with $J^{PC}=1^{--}$ ($^{3}D_{1}$), for all nonets with $n=1$ and $J < 2$ there are corresponding resonances in the PDG (with the interesting exception of the missing orbitally excited vector state $\phi(?)$, with an expected mass at about 1.9 GeV, see Ref. \cite{Piotrowska:2017rgt}). 

Moving to $J \geq 2$, one recognizes the well established nonet of tensor mesons 
with $J^{PC}=2^{++}$ ($^{3}P_{2}$), see e.g Refs. \cite{Burakovsky:1997ci,Giacosa:2005bw,Rodas:2021tyb}, and the quite well-known nonet of pseudotensor mesons with $J^{PC}=2^{-+}$ ($^{1}D_{2}$), even if, for the latter, some questions are still open \cite{Koenigstein:2016tjw,Shastry:2021asu}.
The nonet of mesons with $J^{PC}=3^{--}$ ($^{3}D_{3}$)
is also well known, as the recent detailed report of Ref. \cite{Jafarzade:2021vhh} shows.

On the other hand, as it is visible from Table \ref{mesonlist}, the nonet of the so-called axial-tensor mesons with $J^{PC}=2^{--}$ ($^{3}D_{2}$) is extremely poorly known. Only the kaonic member $K_2(1820)$ (with eventual mixing with $K_2(1770)$) is listed. The isovector member $\rho_2$ and the isotensor ones $\omega_2$ and $\phi_2$ were not yet discovered. One may actually regard this nonet as a sort of ``missing multiplet''. It is then evident that it deserves further theoretical and experimental investigations. 

One of the main goals of the present paper is a phenomenological study of masses, strong, and radiative decays of tensor and, most importantly, of axial-tensor mesons. In the latter case, predictions for the missing states are delivered.  For our purposes, we shall use a well defined chiral model of QCD, called extended Linear Sigma Model (eLSM), that was previously successfully applied to pseudoscalar mesons and vector mesons and their chiral partners, the scalar and the axial-vector mesons \cite{Parganlija:2012fy} (see the first 4 entries of Table \ref{mesonlist}). Within this context, the linear realization of chiral symmetry, together with its explicit and spontaneous breaking patterns, allows to study  chiral partners (that is, $q\bar{q}$ mesons with the same $J$ related by chiral transformations) on the same footing.  Applications of the eLSM to the excited (pseudo)scalar mesons \cite{Parganlija:2016yxq,Janowski:2014ppa}, to pseudovector and orbitally excited vector mesons \cite{Giacosa:2016hrm} and more recently to hybrid mesons \cite{Eshraim:2020ucw} were undertaken. 

It then seems quite natural to apply the eLSM to the tensor sector, thus describing in a chiral framework both tensor and axial-tensor mesons. Since masses and decays of the ground-state tensor mesons are well-established, the tensor mesons allow for the determination of the (additional) parameters of the model and for a test of its validity in the energy range between 1-1.6 GeV. As we shall see, the overall agreement of theory with experiments is rather good and therefore confirms the predominant $q \bar{q}$ nature of the ground-state tensor mesons. 
[For different interpretations of tensor mesons and some other states listed in Table \ref{mesonlist}, see e.g. Refs. \cite{Molina:2008jw,Geng:2010kma,Roca:2012rx}].

Next, we move to the yet mostly unknown axial-tensor mesons. The estimate of masses for the nonet members is possible thanks to the tensor states and to the known axial-tensor kaonic state $K_{2}^{*}(1820)$. 
The decays of axial-tensor mesons into a vector-pseudoscalar pair turns out to be quite large and dominant.  Also the decays into tensor-pseudoscalar and axial-vector-pseudoscalar are sizable and represent interesting decay modes for the future search of the missing states of this nonet. 

\begin{table}[h]
\begin{tabular}{|l|l|l|l|l|l|l|l|}
\hline
$n^{2S+1}L_J$  &  $J^{PC}$ & \begin{tabular}[c]{@{}l@{}}I=1\\ $u\overline{d}$, $d\overline{u}$\\$ \frac{d\overline{d}-u\overline{u}}{\sqrt{2}}$\end{tabular} & \begin{tabular}[c]{@{}l@{}}I=1/2\\ $u\overline{s}$, $d\overline{s}$\\$s\overline{d}$, $s\overline{u}$\end{tabular} & \begin{tabular}[c]{@{}l@{}}I=0\\ $\approx \frac{u\overline{u}+d\overline{d}}{\sqrt{2}}$\end{tabular} & \begin{tabular}[c]{@{}l@{}}I=0\\  $\approx s\overline{s}$\end{tabular} & \begin{tabular}[c]{@{}l@{}} Meson names\end{tabular} & \multirow{2}{*}{\begin{tabular}[c]{@{}l@{}} Chiral \\ Partners \end{tabular}}   \\ \hline
 $1^1 S_0$ & $0^{-+}$ &    $\pi$ & $K$  & $\eta(547)$  & $\eta^\prime(958)$ &    Pseudoscalar  & \multirow{2}{*}{\begin{tabular}[c]{@{}l@{}} $J=0$ \end{tabular}}            \\ \cline{1-7}                                 
  $1^3 P_0$ & $0^{++}$   &    $a_0(1450)$ & $K_0^\star(1430)$  & $f_0(1370)$ & $f_0(1500)/f_0(1710)$ & Scalar   &                                                       \\ \hline
$1^3 S_1$ & $1^{--}$  &   $\rho(770)$  & $K^\star(892)$ & $\omega(782)$ & $\phi(1020)$  & Vector    & \multirow{2}{*}{\begin{tabular}[c]{@{}l@{}} $J=1$ \end{tabular}}            \\ \cline{1-7}                                                        
 $1^3 P_1$ &  $1^{++}$  &  $a_1(1260)$  & $K_{1A}$   & $f_1(1285)$  & $f_1^\prime(1420)$ & Axial-vector   & \\ \hline
  $1^1 P_1$ &  $1^{+-}$   & $b_1(1235)$ & $K_{1 B}$ &$h_1(1170)$ &   $h_1(1415)$ & Pseudovector   & \multirow{2}{*}{\begin{tabular}[c]{@{}l@{}} $J=1^\star$ \end{tabular}}            \\ \cline{1-7}  
  
   $1^3 D_1$ & $1^{--}$  &   $\rho(1700)$  & $K^\star(1680)$ & $\omega(1650)$ & $\phi(???)$  & Excited-vector    &  
     \\ \hline
  $1^3 P_2$ & $2^{++}$  &   $a_2(1320)$ & $K_2^\star(1430)$ & $f_2(1270)$ & $f_2^\prime(1525)$   & Tensor    & \multirow{2}{*}{\begin{tabular}[c]{@{}l@{}} $J=2$ \end{tabular}}            \\ \cline{1-7}  
  
   $1^3 D_2$ & $2^{--}$  & $\rho_2(???)$  & $K_2(1820)$  & $\omega_2(???)$     & $\phi_2(???)$  &  Axial-tensor    &  
     \\ \hline
  $1^1 D_2$ & $2^{-+}$  &   $\pi_2(1670)$ & $K_2(1770)$ & $\eta_2(1645)$ & $\eta_2(1870)$   & Pseudotensor   &                                                      \\ \hline
     $1^3 D_3$ & $3^{--}$  & $\rho_3(1690)$  & $K_3^\star(1780)$  & $\omega_3(1670)$ & $\phi_3(1850)$ &  $J=3$ - Tensor     & 
     \\ \hline 
\end{tabular}\label{mesonlist}
\caption{List of conventional $q\bar{q}$ mesons following the quark model review of Ref. \cite{Zyla:2020zbs} up to $J=3$  and for the lowest radial excitation. The columns refer to different values of the isospin $I$. The first eight nonets are grouped into four chiral partners.  The kaonic states $K_{1,A}$ and $K_{1,B}$ mix and give rise to $K_1(1270)$ and $K_1(1400)$. Note, the entries for which no assignment is proposed are the orbitally excited vector $\phi(???)$ \cite{Piotrowska:2017rgt} as well as three axial-tensor states $\rho_2(???) \equiv \rho_2$, $\omega_2(???) \equiv \omega_2$ and $\phi_2(???) \equiv \phi_2$, which represent one of the main subject of this work.}
\end{table}

Light mesons (both conventional and exotic) are in the centre of various experimental searches, such as the BESIII \cite{Mezzadri:2015lrw,Marcello:2016gcn}, CLAS12 \cite{Rizzo:2016idq}, COMPASS \cite{Ryabchikov:2019rgx}, GlueX \cite{Ghoul:2015ifw,Zihlmann:2010zz,Proceedings:2014joa} experiments as well as the planned PANDA experiment \cite{Lutz:2009ff}. In particular, photoproduction of tensor mesons represents an active experimental \cite{CLAS:2020ngl} and phenomenological \cite{Mathieu:2020zpm,Nys:2018vck,Bibrzycki:2013pja} topic. It is well possible that the missing axial-tensor mesons discussed in this work can be discovered in the  future.

This work is organized as follows: in Sec. \ref{sec:themodel} we introduce the fields and the Lagrangian terms describing masses and decays of (axial-)tensor states; in Sec. \ref{sec:Results} we show the corresponding results. Finally, in Sec. \ref{sec:Conclusion} our conclusions are outlined.


\section{Chiral model for spin-2 mesons}

\label{sec:themodel}
In this Section, we first recall the form of flavour and chiral multiplets and then we present the novel chiral Lagrangian terms that involve (axial-)tensor mesons.

For conventional $q\bar{q}$ pairs, the spin $\vec{S}$ can take the value $S\in\{0\,,1\}$, while the angular momentum $\vec{L}$ any value $L=0,1,2,...$.  
As a consequence, the parity eigenvalue $P=(-1)^{L+1}$ and the charge conjugation eigenvalue $C=(-1)^{L+S}$ follow. 
The spin and angular momentum couple to the total angular momentum $\vec{J}=\vec{L}+\vec{S}$.

For $S=0$, the $q\bar{q}$ state reads $\ket{S=0}$= $\frac{1}{\sqrt{2}}\ket{\uparrow_q\downarrow_{\overline{q}}-\downarrow_q\uparrow_{\overline{q}}}$, while for $S=1$ is
$ \ket{S=1}$ = \{ $\ket{\uparrow_q\uparrow_{\overline{q}}}$\, , $\frac{1}{\sqrt{2}}\ket{\uparrow_q\downarrow_{\overline{q}}+\downarrow_q\uparrow_{\overline{q}}}$ , $\ket{\downarrow_q\downarrow_{\overline{q}}}$ 
\}.
Thus, for $S=0$ 
the possible quantum numbers for conventional $q\bar{q}$ mesons are $J^{PC}\in\Big\{\textbf{even}^{-+},\textbf{odd}^{+-}\Big\}$, while for $S=1$ they are $J^{PC}\in\Big\{\{0,1,2, ...\}^{++},\{1,2,...\}^{--} \Big\}$. [Mesons with
$J^{PC}\in\Big\{\textbf{even}^{+-},\textbf{odd}^{-+}\Big\}$ as well as $0^{--}$ are considered exotic.]

In this work, we concentrate on (pseudo)scalar, (axial-)vector, as well as (axial-)tensor multiplets. These fields enter into appropriate chirally invariant Lagrangian terms that describes the masses and the decays of (axial-)tensor mesons. 

\subsection{Mesonic nonets}
 The ground-state pseudoscalar mesons with $J^{PC}=0^{-+}$ ($^{1}S_{0}$) comprise 3 pions, four kaons, the $\eta(547)$ and the $\eta^\prime(958)$. 
 They can be collected into the following nonet with light-quark elements $P_{ij}\equiv2^{-1/2}\bar{q}_{j}i\gamma^{5}q_{i}$:
\begin{equation}
P=\frac{1}{\sqrt{2}}%
\begin{pmatrix}
\frac{\eta_{N}+\pi^{0}}{\sqrt{2}} & \pi^{+} & K^{+}\\
\pi^{-} & \frac{\eta_{N}-\pi^{0}}{\sqrt{2}} & K^{0}\\
K^{-} & \bar{K}^{0} & \eta_{S}%
\end{pmatrix}
\,\text{,}
\label{eq:pseudoscalar_nonet}%
\end{equation}
where $\eta_{N}\equiv\sqrt{1/2}\,(\bar{u}u+\bar{d}d)$ stands for the
purely non-strange state and $\eta_{S}\equiv\bar{s}s$ for the purely
strange one. 
The isoscalar physical fields emerge upon mixing
reported in e.g. \ Ref. \cite{Kloe2} as:
\begin{equation}
\left(
\begin{array}
[c]{c}%
\eta(547)\\
\eta^{\prime}\equiv\eta(958)
\end{array}
\right)  =\left(
\begin{array}
[c]{cc}%
\cos\beta_{P} & \sin\beta_{P}\\
-\sin\beta_{P} & \cos\beta_{P}%
\end{array}
\right)  \left(
\begin{array}
[c]{c}%
\eta_{N}\\
\eta_{S}%
\end{array}
\right)  \text{ ,}
\end{equation}
with a large mixing angle $\beta_{P}=-43.4^{\circ}$. This sizable admixture is a consequence of the so-called $U_{A}(1)$ axial anomaly \cite{Feldmann:1998vh,tHooft:1986ooh}. Namely, pseudoscalar mesons belong to a so-called  ``heterochiral'' multiplet \cite{Giacosa:2017pos}, see below.

 

Scalar mesons with $J^{PC}=0^{++}$ (following from $L=S=1$, thus $^{3}P_{0}$) are the chiral partners of the pseudoscalar ones. The discussion about their assignment is still ongoing and there is not yet a general agreement. Nevertheless, the set of resonances  \{$a_{0}(1450),$ $K_{0}^{\ast}(1430),$ $f_{0}(1370),$
$f_{0}(1500)/f_{0}(1710)$\} reported in Table \ref{mesonlist} is favoured (here, also the scalar glueball is expected to enter with $f_0(1710)$ being a good candidate). 
Having the quantum numbers of the vacuum, the isoscalar members of this nonet may condense, see Sec. \ref{subsec:chiral}.
By introducing the currents $S_{ij}\equiv2^{-1/2}\bar{q}_{j}q_{i}$, the nonet of scalar states can be written as 
\begin{equation}
S=\frac{1}{\sqrt{2}}%
\begin{pmatrix}
\frac{\sigma_{N}+a_0^{0}}{\sqrt{2}} & a_0^{+} & K_0^{\star +}\\
a_0^{-} & \frac{\sigma_{N}-a_0^{0}}{\sqrt{2}} & K_0^{\star 0}\\
K_0^{\star -} & \bar{K}_0^{\star 0} & \sigma_{S}%
\end{pmatrix}
\,\text{.}
\label{eq:scalar_nonet}%
\end{equation}
The isoscalar sector cannot be described by a simple $2 \times 2$ mixing because (at least) the scalar glueball should be taken into account, thus implying a more complicated  $3 \times 3$ mixing patter leading to the resonances $f_0(1370)$, $f_0(1500)$, and $f_0(1710)$, e.g. Refs. \cite{Llanes-Estrada:2021evz,Janowski:2014ppa,Cheng:2006hu,Giacosa:2005zt,Amsler:1995td,Amsler:1995tu}. Yet, these fields do not appear as final states of decays of (axial-)tensor mesons, therefore its detailed treatment of the scalar-isoscalar mixing is not required in this work.

The third entry in Table \ref{mesonlist} refers to the $J^{PC}=1^{--}$ ($^{3}S_{1}$) nonet with $L=0$ and $S=1$. 
These are the very well known vector
states \{$\rho(770),$ $K^{\ast}(892),$ $\omega(782),$ $\phi(1020)$\}. The matrix $V^{\mu}$ with elements
$V_{ij}^{\mu}=2^{-1/2}\bar{q}_{j}\gamma^{\mu}q_{i}$ has the form
\begin{equation}
V^{\mu}=\frac{1}{\sqrt{2}}%
\begin{pmatrix}
\frac{\omega_{N}^{\mu}+\rho^{0\mu}}{\sqrt{2}} & \rho^{+\mu} &
K^{\ast+\mu}\\
\rho^{-\mu} & \frac{\omega_{N}^{\mu}-\rho^{0\mu}}{\sqrt{2}} & K^{\ast0\mu}\\
K^{\ast-\mu} & \bar{K}^{\ast0\mu} & \omega_{S}^{\mu}%
\end{pmatrix}
\,, \label{eq:vector_nonet}%
\end{equation}
where $\omega_{N}$ and $\omega_{S}$ are purely non-strange and strange,
respectively. Similar to the pseudoscalar case, the physical fields arise upon mixing
\begin{equation}
\left(
\begin{array}
[c]{c}%
\omega(782)\\
\phi(1020)
\end{array}
\right)  =\left(
\begin{array}
[c]{cc}%
\cos\beta_{V} & \sin\beta_{V}\\
-\sin\beta_{V} & \cos\beta_{V}%
\end{array}
\right)  \left(
\begin{array}
[c]{c}%
\omega_{N}\\
\omega_{S}%
\end{array}
\right)  \text{ ,}%
\end{equation}
where the small isoscalar-vector mixing angle $\beta_{V}=-3.9^{\circ}$ is
taken from the PDG \cite{Zyla:2020zbs}. Thus, the physical states $\omega$ and $\phi$
are dominated by non-strange and strange components, respectively. This is in
agreement with the so-called ``homochiral'' nature of these states \cite{Giacosa:2017pos}, see also below.


The chiral partners of the vector mesons \cite{Hatanaka:2008gu} are realized for the quantum numbers $L=S=1$ (just as for the scalar mesons discussed previously) coupled to  $J^{PC}=1^{++}$ ($^{3}P_{1}$): this is the so-called axial-vector meson nonet $A_{1}$, that arises
from the
microscopic current $A_{1,ij}^{\mu}\equiv2^{-1/2}\bar{q}_{j}
\gamma^{5}\gamma^{\mu}  q_{i}$:
\begin{equation}
A_{1}^{\mu}=\frac{1}{\sqrt{2}}%
\begin{pmatrix}
\frac{f_{1,N}^{\mu}+a_{1}^{0\mu}}{\sqrt{2}} & a_{1}^{+\mu} & K_{1A}^{+\mu}\\
a_{1}^{-\mu} & \frac{f_{1,N}^{\mu}-a_{1}^{0\mu}}{\sqrt{2}} & K_{1A}^{0\mu}\\
K_{1A}^{-\mu} & \bar{K}_{1A}^{0\mu} & f_{1,S}^{\mu}%
\end{pmatrix}
\text{ }\,. \label{eq:avector_nonet}%
\end{equation}
The isoscalar sector reads
\begin{equation}
\left(
\begin{array}
[c]{c}%
f_{1}(1285)\\
f_{1}(1420)
\end{array}
\right)  =\left(
\begin{array}
[c]{cc}%
\cos\beta_{A_{1}} & \sin\beta_{A_{1}}\\
-\sin\beta_{A_{1}} & \cos\beta_{A_{1}}%
\end{array}
\right)  \left(
\begin{array}
[c]{c}%
f_{1,N}^{\mu}\\
f_{1,S}^{\mu}%
\end{array}
\right)  \text{ ,}%
\end{equation}
where the mixing angle $\beta_{A_{1}}$ is expected to be small because of the homochiral nature of the multiplet \cite{Giacosa:2017pos}, see also Ref. \cite{Divotgey:2013jba}. Here, we shall set this angle to zero for simplicity.  

An important digression is needed for the kaonic axial-vector mesons. Since these states are not
eigenstates of $C$, the state $K_{1,A}$ belonging to the $J^{PC}=1^{++}$ ($^{3}P_{1}$) axial-vector nonet and $K_{1,B}$ belonging to the $J^{PC}=1^{+-}$ ($^{1}P_{1}$) pseudovector nonet mix:
\begin{equation}
\left(
\begin{array}
[c]{c}%
K_{1}(1270)\\
K_{1}(1400)
\end{array}
\right)  ^{\mu}=\left(
\begin{array}
[c]{cc}%
\cos\varphi_{K} & -i\sin\varphi_{K}\\
-i\sin\varphi_{K} & \cos\varphi_{K}%
\end{array}
\right)  \left(
\begin{array}
[c]{c}%
K_{1,A}\\
K_{1,B}%
\end{array}
\right)  ^{\mu}\text{ .} \label{mixk1abfields}%
\end{equation}


For a detailed study on the $K_{1}(1270)$%
/$K_{1}(1400)$ system using an approach similar to the one used in this work, see Ref. \cite{Divotgey:2013jba}.  The numerical value $\varphi_{K}=\left(  56.4\pm4.3\right)  ^{\circ}$ \cite{Divotgey:2013jba,Hatanaka:2008gu} implies that 
$K_{1}(1270)$ is predominantly $K_{1,B}$ and $K_{1}(1400)$ is predominantly $K_{1,A}$ (but the mixing is large). The study of this mixing angle via semileptonic decays
can be found in e.g. Ref. \cite{Cheng:2017pcq}.

The next (fifth and sixth) entries of Table \ref{mesonlist} are the pseudovector (from which the $K_{1,B}$ above stems) and their chiral partners, the (orbitally) excited vector mesons. One can build similar nonets (see Refs. \cite{Giacosa:2016hrm,Eshraim:2020ucw}), yet they are not part of the decay products of the (axial-)tensor mesons studied in this paper, therefore we omit them in the following.

Finally, we present the two tensor ($J=2$) meson nonets (seventh and eighth rows of Table \ref{mesonlist}), which constitute the main subject of this work.

The well-known $J^{PC}=2^{++}$ ($^{3}P_{2}$) tensor states with elements $T_{ij}^{\mu\nu}=2^{-1/2}\bar{q}
_{j}(i\gamma^{\mu}\partial^\nu+\cdots)q_{i}$ 
form an almost ideal nonet of quark-antiquark states:
\begin{equation}
T^{\mu\nu}=\frac{1}{\sqrt{2}}%
\begin{pmatrix}
\frac{f_{2,N}^{\mu\nu}+a_{2}^{0\mu\nu}}{\sqrt{2}} & a_{2}^{+\mu\nu} &
K_{2}^{\ast+\mu\nu}\\
a_{2}^{-\mu\nu} & \frac{f_{2,N}^{\mu\nu}-a_{2}^{0\mu\nu}}{\sqrt{2}} &
K_{2}^{\ast0\mu\nu}\\
K_{2}^{\ast-\mu\nu} & \bar{K}_{2}^{\ast0\mu\nu} & f_{2,S}^{\mu\nu}%
\end{pmatrix}
\,. \label{eq:tensor_nonet}%
\end{equation}
The physical isoscalar-tensor states are
\begin{equation}\label{tens-mix}
\left(
\begin{array}
[c]{c}%
f_{2}(1270)\\
f_{2}^{\prime}(1525)
\end{array}
\right)  =\left(
\begin{array}
[c]{cc}%
\cos\beta_{T} & \sin\beta_{T}\\
-\sin\beta_{T} & \cos\beta_{T}%
\end{array}
\right)  \left(
\begin{array}
[c]{c}%
f_{2,N}\\
f_{2,S}%
\end{array}
\right)  \text{  ,}%
\end{equation}
where $\beta_{T} \simeq 5.7^{\circ}$ is the small mixing angle reported in the
PDG, in agreement with the fact that tensor mesons belong to a homochiral
multiplet, just as (axial-)vector states. Later on, we shall re-calculate the mixing angle $\beta_T$ by a fit to data on decays.
The decays of tensor mesons were studied in detail in Refs.
\cite{Giacosa:2005bw,Burakovsky:1997ci} and refs. therein, and fit  well into the standard $q\bar{q}$ picture \cite{isgur1985}. 
This point is corroborated by the results of this work, see the next Section.



The axial-tensor nonet with $J^{PC}=2^{--}$ ($^{3}D_{2}$) with currents $2^{-1/2}\bar{q}_{j}\left(  \gamma^{5}%
\gamma^{\mu}\partial^{\nu}+...\right)  q_{i}$, builds the chiral partners of the previously introduced tensor mesons. This nonet is however poorly known: at present, only the kaonic states $K_{2}(1820)$ and $K_{2}(1770)$ have been measured, the heavier one being considered as predominately axial-tensor and the lighter mainly pseudotensor, see Table \ref{mesonlist} (as for the axial-vector and pseudovector kaons, mixing is possible, but at present not calculable due to lack of experimental data).  For some recent theoretical works devoted to the vacuum properties and thermal properties of these resonances, see Refs. \cite{Guo:2019wpx,Abreu:2020wio}  and Refs. \cite{Turkan:2019vnc,Sungu:2020azn} respectively. 
The matrix for this nonet reads: 
\begin{align}
    	A_2^{\mu\nu} & =\frac{1}{\sqrt{2}}
		\begin{pmatrix}
			\frac{\omega_{2,N}^{\mu\nu} + \rho_{2}^{0\mu\nu}}{\sqrt{2}} & \rho_{2}^{+\mu\nu} &	K_{2}^{+\mu\nu} \\
			\rho_{2}^{-\mu\nu} & 	\frac{\omega_{2,N}^{\mu\nu} -\rho_{2}^{0\mu\nu}}{\sqrt{2}}&	K_{2}^{0\mu\nu}  \\
			K_{2}^{-\mu\nu}  & \bar{K}_{2}^{0\mu\nu} & \omega_{2,S}^{\mu\nu}
		\end{pmatrix}\,.
		\label{eq:tensor3_nonet}
\end{align}
The isoscalar mixing is expected to be small in view of the homochiral nature of the corresponding chiral multiplet in which this
nonet is embedded \cite{Giacosa:2017pos} (see also below).  Thus, for the isoscalar members: $\omega_{2,N} \simeq \omega_2$ and $\omega_{2,S} \simeq \phi_2$.

The transformation rules under parity, charge conjugation, and flavor transformations for all the relevant nonets of this work ($P, S, V, A_1, T, A_2$) are summarized in Table \ref{tab:transformations}. 

\begin{table}[h] 
		\centering
		\renewcommand{\arraystretch}{1.1}
		\begin{tabular}[c]{|c|c|c|c|}
			\hline
			Nonet & Parity $(P)$ & Charge conjugation $(C)$ & Flavour $(U_{V}(3))$ \\
			\hline \hline
				$0^{-+}=P$ & $-P(t,-\vec{x})$ & $P^{t}$ & $U P U^{\dagger}$ \\
			\hline
			$0^{++}=S$ & $S(t,-\vec{x})$ & $S^{t}$ & $U S U^{\dagger}$ \\
			\hline
						$1^{--}=V^{\mu}$ & $V_{\mu}(t,-\vec{x})$ & $-(V^{\mu})^{t}$ & $U V^{\mu}U^{\dagger}$ \\
			\hline
				$1^{++}=A_1^{\mu}$ & $-A_{1\,\mu}(t,-\vec{x})$ & $(A_1^{\mu})^{t}$ & $U A_1^{\mu}U^{\dagger}$ \\
			\hline
							$2^{++}=T^{\mu\nu}$ & $T_{\mu\nu}(t,-\vec{x})$ & $(T^{\mu\nu})^{t}$ & $U T^{\mu\nu} U^{\dagger}$ \\
			\hline
			$2^{--}=A_{2}^{\mu\nu}$ & $-A_{2\,\mu\nu}(t,-\vec{x})$ & $-(A_2^{\mu\nu})^{t}$ & $U A_2^{\mu\nu} U^{\dagger}$ \\
			\hline
		\end{tabular}
			\caption{Mesonic nonet transformations under P, C and $U_V(3)$.}
		\label{tab:transformations}
		\end{table}

Finally, we introduce the chiral multiplets  which contain chiral partners and have simple transformations under the chiral group $U_R(3) \times U_L(3)$. 

The pseudoscalar and scalar nonets build the matrix 
\begin{align}
   \Phi = S+iP \text{ ,{}}
\end{align}
whose ``heterochiral'' transformation can be found in Table \ref{tab:chiral-transformations}. As a consequence, the strange-nonstrange mixing angle can be large, as it is the case for the pseudoscalar mesons (the scalar one is not easy to determine because one has a three-state mixing problem) and possibly for the pseudotesnor mesons \cite{Giacosa:2017pos,Koenigstein:2016tjw}.


Vector and axial-vector nonets and similarly tensor and axial-tensor nonets build the following left-handed and right-handed objects:
\begin{align}
     L^{\mu}:= V^{\mu}+A_1^{\mu} \ \text{ , }  R^{\mu}:= V^{\mu}-A_1^{\mu} \text{ ,}
\end{align}
\begin{align}
    \mathbf{L}^{\mu\nu}:=T^{\mu \nu}+A_2^{\mu \nu} \text{ , }  \mathbf{R}^{\mu\nu}:=T^{\mu \nu}-A_2^{\mu \nu}\text{ .}
\end{align}
Their homochiral transformations are also listed in Table \ref{tab:chiral-transformations}. One expects small isoscalar mixing angles in these four nonets \cite{Giacosa:2017pos}.

The fields listed in Table \ref{tab:chiral-transformations} are the needed ingredients to write down chiral Lagrangian terms, as we shall show in the next subsection.

  	\begin{table}[h]
   	\centering
		\renewcommand{\arraystretch}{1.1}
		\begin{tabular}[c]{|c|c|c|c|}
			\hline
			Nonet & Parity $(P)$ & Charge conjugation $(C)$ & $U_R(3) \times U_L(3)$ \\
			\hline \hline
			$\Phi(t,\vec{x})$ & $\Phi^{\dagger}(t,-\vec{x})$ & $\Phi^{t}(t,\vec{x})$ & $U_L\Phi U^{\dagger}_R$ \\
			\hline
			$R^{\mu}(t,\vec{x})$ & $L_{\mu}(t,-\vec{x})$ & $-(L^{\mu }(t,\vec{x}))^{t}$ & $U_R R^{\mu}U_R^{\dagger}$ \\
			\hline
			$L^{\mu}(t,\vec{x})$ & $R_{\mu}(t,-\vec{x})$ & $-(R^{\mu }(t,\vec{x}))^{t}$ & $U_L L^{\mu}U_L^{\dagger}$ \\
			\hline
			$\mathbf{R}^{\mu\nu}(t,\vec{x})$ & $\mathbf{L}_{\mu\nu}(t,-\vec{x})$ & $(\mathbf{L}^{\mu\nu}(t,\vec{x}))^{t}$ & $U_{R}\mathbf{R}^{\mu\nu}U^{\dagger}_{R}$ \\
			\hline
			$\mathbf{L}^{\mu\nu}(t,\vec{x})$ & $\mathbf{R}_{\mu\nu}(t,-\vec{x})$ & $(\mathbf{R}^{\mu\nu}(t,\vec{x}))^{t}$ & $U_{L}\mathbf{L}^{\mu\nu}U^{\dagger}_{L}$ \\
			\hline
		\end{tabular}
		\caption{Transformations of the chiral multiplets under P, C, and $U_R(3) \times U_L(3)$.}\label{tab:chiral-transformations}
					\end{table}



\subsection{Chiral invariant Lagrangian terms}
\label{subsec:chiral}
The standard eLSM Lagrangian is a function of (pseudo)scalar and (axial-)vector mesons as well as a dilaton (or scalar glueball) field $G$, see Refs. \cite{Parganlija:2012fy,Janowski:2014ppa,Eshraim:2020ucw}:
\begin{align}\nonumber
\mathcal{L}_{\text{eLSM}} = \mathcal{L}_{\text{dil}}+ \text{Tr}\Big[\Big(D_{\mu}\Phi\Big)^\dagger\Big(D_{\mu}\Phi\Big)\Big]-m_{0}^2\Big(\frac{G}{G_0}\Big)^2\text{Tr}\Big[\Phi^\dagger\Phi\Big]-\frac{1}{4}\text{Tr}\Big[\Big(L_{\mu\nu}^2+R_{\mu\nu}^2\Big)\Big]\\
+\text{Tr}\Big[\Big(\frac{m_{\text{vec}}^2G^2}{2G_0^2}+\Delta\Big)\Big(L_{\mu}^2+R_{\mu}^2\Big)\Big]+\cdots
\text{ ,}
\end{align}
where $\mathcal{L}_{\text{dil}}$ is the dilaton Lagrangian for the field $G$ \cite{Migdal:1982jp,Salomone:1980sp} and $G_0$ is the corresponding vacuum expectation value (v.e.v.). [Note, the dilaton field $G$ will not be relevant but is formally important for dilatation invariance, but will not appear as a decay product.]  
The covariant derivative entering in the previous equation reads $D_{\mu}\Phi:=\partial_\mu\Phi-ig_1\big(L_\mu\Phi-\Phi R_\mu\big)$.
The matrix $\Delta$ breaks flavour symmetry by including a different mass for mesons carrying the strange quark:
\begin{equation}
   \Delta = \begin{pmatrix}
\delta_N & 0 &
0\\
0 & \delta_N & 0\\
0 & 0 & \delta_S%
\end{pmatrix}\;\; \text{where}\;\; \delta_N \thicksim m_u^2 \ \text{ ,} \quad \delta_S \thicksim m_s^2
\text{ .}
\end{equation}
 The eLSM is based on the proper treatment of dilatation and chiral invariances and their explicit and spontaneous breaking patterns. [For other chiral models that make use of dilatation invariance, see e.g. Refs. \cite{Ko:1994en,Carter:1995zi} and refs. therein]. 
The results for masses and decays of (pseudo)scalar and (axial-)vector mesons can be found in the afore mentioned Refs. \cite{Parganlija:2012fy,Janowski:2014ppa}. For our present work, we need to recall some of its basic features.

(i) The scalar-isoscalar fields of the model acquire nonzero v.e.v. The dilaton field $G$ condenses to a certain value $G_0$ (whose numerical value is not needed here), which is related to the breaking of dilatation invariance.  Similarly, $q \bar{q} $ scalar-isoscalar fields develop a v.e.v. as a consequence of spontaneous symmetry breaking (SSB), which in simple terms is due to the Mexican hat form of the effective potential when $m_{0}^{2} <0$. 
As a consequence, shifts are  necessary: 
\begin{align}
   G\rightarrow G+ G_0 \text{ ,  } S\rightarrow S+ \Phi_0 \qquad \text{with}\qquad \Phi_0:=\frac{1}{\sqrt{2}}\begin{pmatrix}
\frac{\phi_N}{\sqrt{2}} & 0 &
0\\
0 & \frac{\phi_N}{\sqrt{2}} & 0\\
0 & 0 & \phi_S%
\end{pmatrix}
\text{ .}
\label{vev}
\end{align}
where the numerical values are reported in Table \ref{constants}.

(ii) The axial-vector matrix must be shifted because SSB induces a mixing of scalar and axial-vector fields,  which is proportional to the parameter $g_1= \frac{m_{a_1}}{Z_{\pi}f_{\pi}}\sqrt{1-\frac{1}{Z_{\pi}^2}} \simeq 5.8$ \cite{Parganlija:2012fy,Parganlija:2016yxq}. This operation amounts to:
\begin{align}
        A_{1\,\mu}\rightarrow A_{1\,\mu}+\partial_{\mu}\mathcal{P} ,\,\,  \mathcal{P} := \frac{1}{\sqrt{2}}
		\begin{pmatrix}
			\frac{Z_{\pi}w_{\pi}(\eta_N + \pi^0)}{\sqrt{2}} & Z_{\pi}w_{\pi}\pi^+ & Z_Kw_K K^+ \\
			Z_{\pi}w_{\pi}\pi^- & \frac{Z_{\pi}w_{\pi}(\eta_N - \pi^0)}{\sqrt{2}} &  Z_Kw_K K^0 \\
			 Z_Kw_K K^- &  Z_Kw_K \bar{K}^0 & Z_{\eta_S}w_{\eta_S}\eta_S
		\end{pmatrix}
		\label{shiftedP}
		\text{ ,}
    \end{align}
where the constant terms are also given in Table \ref{constants}. The matrix $\mathcal{P}$ contains the pseudoscalar fields, just as the matrix $P$ of Eq. (\ref{eq:pseudoscalar_nonet}), but proper renormalization factors have been included. 
The shifts in Eqs. (\ref{vev}) and (\ref{shiftedP}) shall also be applied to the novel Lagrangian terms introduced below. 
\begin{table}[h]
\centering
\renewcommand{\arraystretch}{1.2} 
\begin{tabular}{|c|c|c|}
\hline
\text{Parameters} &\text{Expressions}  & \text{ Numerical values }    \cite{Parganlija:2012fy} \\ \hline
$\,\;Z_{\pi}=Z_{\eta_N}$ &  $\frac{m_{a_1}}{\sqrt{m_{a_1}^2- g_1^2 \phi_S^2}}$  & $1.709$    \\ \hline
$\,\;w_{\pi}=w_{\eta_N}$ & $\frac{g_1 \phi_N}{m_{a_1}^2}$  & $0.683\,\text{GeV}^{-1}$  \\ \hline
$\,\;Z_{K}$ & $\frac{2m_{K_{1A}}}{\sqrt{4m_{K_{1A}}^2- g_1^2 (\phi_N+\sqrt{2}\phi_S)^2}}$  & $1.604$  \\ \hline
$\,\;w_K$ & $\frac{g_1 (\phi_N+\sqrt{2}\phi_S)}{2m_{K_{1,A}}^2}$  & $0.611\,\text{GeV}^{-1}$ \\ \hline
 $\,\;Z_{\eta_S}$ & $\frac{m_{f_{1,S}}}{\sqrt{m_{f_{1,S}}^2-2 g_1^2\phi_S^2}}$ & $1.539$  \\ \hline
$\,\;w_{\eta_S}$ & $\frac{\sqrt{2} g_1 \phi_S}{m_{f_{1,S}}^2}$ & $0.554\,\text{GeV}^{-1}$  \\ \hline
$\,\;\phi_N$ & $ Z_{\pi}f_{\pi}$ & $0.158\, \text{GeV}$\\\hline 
$\,\;\phi_S$  & $\frac{2Z_Kf_K-\phi_{N}}{\sqrt{2}}$ & $0.138\, \text{GeV}$\\ \hline
$\,\;g_1$  & $\frac{m_{a_1}}{Z_{\pi}f_{\pi}}\sqrt{1-\frac{1}{Z_{\pi}^2}}$ & $5.8$\\ \hline
\end{tabular}
\caption{Numerical values of (scalar and axial-vector) shift-related quantities needed in this work. }\label{constants}
\end{table}

Next, we turn to the chirally invariant Lagrangian terms that contain (axial-)tensor mesons. For convenience, we shall split them into four distinct ones.

First, the chiral invariant Lagrangian that generates, among other interactions, the masses of the spin-2 mesons reads
\begin{align}\label{eq:mLag}
    \mathcal{L}_{\text{mass}}=\text{Tr}\Big[\Big(\frac{m_{\text{ten}}^{2}\,G^2}{2\,G_0^2}+\Delta^{\text{ten}}\Big)\Big(\mathbf{L}_{\mu\nu}^2+\mathbf{R}_{\mu\nu}^2\Big)\Big]
   + \frac{h_1^{\text{ten}}}{2}\text{Tr}\Big[\Phi^\dagger \Phi\Big]\text{Tr}\Big[\mathbf{L}^{\mu\nu} \mathbf{L}_{\mu\nu}+ \mathbf{R}^{\mu\nu} \mathbf{R}_{\mu\nu}\Big]  \\\nonumber
   +h_2^{\text{ten}}\text{Tr}\Big[\Phi^\dagger \mathbf{L}^{\mu\nu} \mathbf{L}_{\mu\nu} \Phi+\Phi \mathbf{R}^{\mu\nu} \mathbf{R}_{\mu\nu} \Phi^\dagger\Big]+2 h_3^{\text{ten}}\text{Tr}\Big[\Phi \mathbf{R}^{\mu\nu}\Phi^\dagger \mathbf{L}_{\mu\nu} \Big]
   \text{,}
\end{align}
where $\Delta^{\text{ten}}=\text{diag}\{\delta_{N}^{\text{ten}}\,,\delta_{N}^{\text{ten}}\,,\delta_{S}^{\text{ten}}\}$; one may set the (anyhow small) light-quark contribution $\delta_{N}^{\text{ten}}=0$, since a term proportional to the identity matrix can be reabsorbed into the term proportional to $m_{\text{ten}}$; the element $\delta_S^{\text{ten}}$ 
is then obtained from the mass relation $\delta_S^{\text{ten}}= m_{K_{2}}^2-m_{\textbf{a}_2}^2 \simeq 0.3\,\text{GeV}^2$ using PDG values for the tensor mesons (details in Sec. \ref{shiftedP}).
Besides the term proportional to $\Delta^{\text{ten}}$, all the other terms involve dimensionless coupling constants. 

According to the large-$N_c$ limit, the coupling constants $h_{2}^{\text{ten}}$ and $h_{3}^{\text{ten}}$ scale as $N_c^{-1}$, while $h_{1}^{\text{ten}}$ as $N_c^{-3}$, thus being suppressed. The mass term $m_{\text{ten}}$ is $N_c$-independent (it is a mass term that sets the mass scale of (axial-)tensor mesons), while the condensate $G_0$  scales as $N_c$. As a consequence, the dimensionless coupling $m_{\text{ten}}^{2}/G_0^{2}$ scales as $N_c^{-2}$.

We then turn to other interaction terms that generate the decays of (axial-)tensor mesons.  The first of such terms involves solely left- and right-handed chiral fields:
\begin{align}\label{lag1}
    \mathcal{L}_{g_2^{\text{ten}}}=\frac{g_2^{\text{ten}}}{2}\Big(\text{Tr}\Big[ \mathbf{L}_{\mu\nu}\{ L^{\mu}, L^{\nu}\}\Big]+\text{Tr}\Big[\mathbf{R}_{\mu\nu} \{R^{\mu}, R^{\nu}\} \Big]\Big)  +\frac{g_2^{\prime\,\text{ten}}}{6}\text{Tr}\Big[\mathbf{L}_{\mu\nu}+\mathbf{R}_{\mu\nu}\Big]\text{Tr}\Big[\{L^{\mu}, L^{\nu}\}+\{R^{\mu}, R^{\nu}\}\Big]\
    \text{ ,}
\end{align}
whose coupling constants scales as $g_2^{\text{ten}} \propto N_c^{-1/2}$ (dominant) and $g_2^{\prime \text{ten}} \propto N_c^{-3/2}$ (suppressed).  Further suppressed terms are possible, but are not considered here. The coupling constants have dimension [Energy], thus the Lagrangian can be made dilatation invariant by multiplying by $G/G_0$. (Upon condensation of $G$, the term above is reobtained; this is omitted here, since no relevant additional interactions emerge).  
The Lagrangian $\mathcal{L}_{g_2^{\text{ten}}}$ enables us to compute the following two types of decays:
\begin{enumerate}
    \item $2^{++}\longrightarrow 0^{-+} + 0^{-+}$ ;
     \item $2^{--}\longrightarrow 0^{-+} + 1^{--}$ .
\end{enumerate}
Both decays are extremely relevant in this work, since the pseudoscalar-pseudoscalar mode is the dominant one for the tensor mesons, while the pseudoscalar-vector one is the dominant channel for the axial-tensor states. Eq. (\ref{lag1}) shows that they are related by chiral transformations, and hence depend on the same coupling constant(s). 

The next Lagrangian that couples left-handed and right-handed fields involves derivatives: 
\begin{align}\label{lag2}
    \mathcal{L}_{a^{\text{ten}}}=\frac{a^{\text{ten}}}{2}\text{Tr}\Big[\mathbf{L}_{\mu\nu}\{L^{\mu}_{\beta},L^{\nu\beta}\}+\mathbf{R}_{\mu\nu}\{R^{\mu}_{\beta},R^{\nu\beta}\}\Big]
    +\frac{a^{\prime\,\text{ten}}}{6}\text{Tr}\Big[\mathbf{L}_{\mu\nu}+\mathbf{R}_{\mu\nu}\big]\text{Tr}\Big[\{L^{\mu}_{\beta},L^{\nu\beta}\}+\{R^{\mu}_{\beta},R^{\nu\beta}\}\Big]\,,
\end{align}
where $L^{\mu\nu}:=\partial^{\mu}L^{\nu}-\partial^{\nu}L^{\mu}$ and analogously for  $R^{\mu\nu}$.
The large-$N_c$ scaling behaviors are $a^{\text{ten}} \propto N_c^{-1/2}$ (dominant) and  $a^{\prime \text{ten}} \propto N_c^{-3/2}$ (subdominant). This Lagrangian cannot be made dilatation invariant\footnote{Note, formally one could multiply the Lagrangian of Eq. (\ref{lag2}) by $G_0/G$, but that implies a non-analytic point of the interaction potential for $G \rightarrow 0$; this is why only terms with coupling constants with zero or positive dimension can be made dilatation invariant.} because the coupling constants have dimension [Energy$^{-1}$], thus it is expected to deliver suppressed decays into mesons (as it turns out to be the case). Yet, it is relevant because, after an appropriate inclusion of the photon field via vector meson dominance (VMD), it describes the decay of the tensor mesons to two photons, one photon and one vector mesons, as well as giving a subleading but important contribution  to the decay into two vector mesons.

The next chiral interaction term reads
\begin{align}\nonumber\label{chiralag-vp}
    \mathcal{L}_{c^{\text{ten}}}=c_1^{\text{ten}}\,\text{Tr} \Big[ \partial^{\mu}\textbf{L}^{\nu\alpha} \tilde{L}_{\mu\nu}\,\partial_{\alpha}\Phi\,\Phi^\dagger-\partial^{\mu}\textbf{R}^{\nu\alpha} \Phi^\dagger\,\partial_{\alpha}\Phi \tilde{R}_{\mu\nu}- \partial^{\mu}\textbf{R}^{\nu\alpha} \tilde{R}_{\mu\nu}\partial_{\alpha}\Phi^\dagger \Phi +\partial^{\mu}\textbf{L}^{\nu\alpha}\Phi \partial_{\alpha} \Phi^\dagger \tilde{L}_{\mu\nu}\Big]\\
   + c_2^{\text{ten}}\,\text{Tr} \Big[ \partial^{\mu}\textbf{L}^{\nu\alpha}\partial_{\alpha}\Phi \tilde{R}_{\mu\nu}\,\,\Phi^\dagger-\partial^{\mu}\textbf{R}^{\nu\alpha} \Phi^\dagger\,\tilde{L}_{\mu\nu}\partial_{\alpha}\Phi - \partial^{\mu}\textbf{R}^{\nu\alpha} \partial_{\alpha}\Phi^\dagger\tilde{L}_{\mu\nu} \Phi +\partial^{\mu}\textbf{L}^{\nu\alpha}\Phi\tilde{R}_{\mu\nu} \partial_{\alpha} \Phi^\dagger \Big]
   \text{ ,}
\end{align}
where $\tilde{L}_{\mu\nu}:=\frac{\varepsilon_{\mu\nu\rho\sigma}}{2}(\partial^{\rho}L^{\sigma}-\partial^{\sigma}L^{\rho})$ and similarly for $\tilde{R}_{\mu\nu}$. This Lagrangian breaks also dilatation invariance as it is typical for terms involving the Levi-Civita tensor.  Both constants $c_1^{\text{ten}}$ and $c_2^{\text{ten}}$ scale as $N_c^{-1/2}$. Large-$N_c$ suppressed terms are possible but are omitted here since present data do not allow to constrain them. 
The relevant decay terms are:
\begin{enumerate}
    \item $2^{++}\longrightarrow 0^{-+} + 1^{--}$;
     \item $2^{--}\longrightarrow 0^{-+} + 1^{++}$.
\end{enumerate}
These decay channels are related by chiral symmetry and for tensor mesons they are quite relevant. Note, the pseudoscalar meson as a decay product is a consequence of the shift of Eq. \eqref{shiftedP}.


The decay of a generic spin-2 state has the form:
\begin{equation}\label{decayform}
\Gamma_{T\rightarrow A+B}^{tl}=\frac{|\vec{k}_{a,b}%
|}{40\,\pi\,m_{t}^{2}}\times|\mathcal{M}|^{2}\times\kappa_{i}\times\Theta
(m_{t}-m_{a}-m_{b})\,,%
\end{equation}
where $m_{t}$ is the mass of the decaying (axial-)tensor particle, $m_{a}$ and
$m_{b}$ are the masses of the decay products, $\Theta(x)$ denotes the
Heaviside step-function, and the modulus of the outgoing particles momentum
has the following analytic expression:
\begin{equation}
|\vec{k}_{a,b}|:=\frac{1}{2\,m_{t}}\sqrt{(m_{t}^{2}-m_{a}^{2}-m_{b}^{2}%
)^{2}-4m_{a}^{2}m_{b}^{2}}\ \text{ .} \label{eq:kf}%
\end{equation}
We obtain the (eventually dimensionful) factors $\kappa_{i}$ entering in Eq. (\ref{decayform}) for the $i$-th
decay channel from the explicit forms of the Lagrangian (the index $i$ runs over the flavor channels, see the following). 
The decay amplitudes
$|\mathcal{M}|^{2}$, derived via Feynman rules under the use of the
polarization vectors (tensors) and their corresponding completeness relations,
are listed in Table \ref{tab:amplitudes}.
\begin{table}[ptb]
\centering
\renewcommand{\arraystretch}{1.1}
\begin{tabular}
[c]{|c|c|}\hline
Decay Mode & $\frac{1}{5}\times|\mathcal{M}|^{2}$\\\hline\hline
$2^{++}\longrightarrow0^{-+}+0^{-+}$ & $g_2^{\text{ten}\,2}\times\frac{2|\vec
{k}_{p^{(1)},p^{(2)}}|^{4}}{15} $\\\hline\hline
$2^{--}\longrightarrow0^{-+}+1^{--}$ & $g_2^{\text{ten}\,2}\times\frac{|\vec
{k}_{v,p}|^2}{15}\Big(5+\frac{2|\vec
{k}_{v,p}|^2}{m_v^2}\Big)$\\\hline\hline
$2^{--}\longrightarrow 0^{-+}+2^{++}$ & $\frac{4 h_3^{\text{ten}\,2}}{45}\Big(45+\frac{4|\vec{k}_{t,p}%
|^4}{m_{t}^4}+\frac{30|\vec{k}_{t,p}%
|^2}{m_{t}^2}\Big)$\\\hline\hline
$2^{++}\longrightarrow0^{-+}+1^{--}$ & $\frac{(c_1^{\text{ten}}+c_2^{\text{ten}})^2\,m_t^2|\vec
{k}_{v,p}|^4}{5}$\\\hline\hline
$2^{--}\longrightarrow0^{-+}+1^{++}$ & $\frac{(c_1^{\text{ten}}-c_2^{\text{ten}})^2\,m_{a_2}^2|\vec
{k}_{a_1,p}|^4}{5}$\\\hline
\end{tabular}
\caption{Decay amplitudes for different decay modes.}%
\label{tab:amplitudes}%
\end{table}
To this end, we recall that one  has to average over all incoming spin-polarizations, sum up all possible outgoing ones, as well as considering the following completeness relations (for further details see e.g. \cite{Koenigstein:2015asa}):
	\begin{align}
		\sum_{\lambda=1}^{3} \epsilon_{\mu}(\lambda ,\, \vec{k})\, \epsilon_{\nu}(\lambda ,\, \vec{k}) & = -G_{\mu\nu}\,, \label{eq:completenessvector}
		\\
		\sum_{\lambda=1}^{5} \epsilon_{\mu\nu} (\lambda ,\, \vec{k})\, \epsilon_{\alpha\beta} (\lambda ,\, \vec{k}) & = -\frac{G_{\mu\nu} G_{\alpha\beta}}{3} + \frac{G_{\mu\alpha} G_{\nu\beta} + G_{\mu\beta} G_{\nu\alpha}}{2} \,, \label{eq:completenesstensor}
	\end{align}
		where
\begin{align}
		G_{\mu\nu} & \equiv \eta_{\mu\nu} - \frac{k_{\mu} k_{\nu}}{m^{2}} \,,\,\,\,\,\, \eta_{\mu\nu}\equiv \begin{pmatrix}
1 & 0 & 0 & 0\\
0 & -1 & 0 & 0\\
0 & 0 & -1 & 0\\
0 & 0 & 0 & -1
\end{pmatrix}
\text{ .}
	\end{align}

\section{Results}
\label{sec:Results}
In this section we present the results for the (axial-)tensor mesons. First, we discuss the masses of the well established tensor mesons and of their poorly known chiral partners, the axial-tensor ones. Then, we turn to the decays, first those of tensor mesons (that are well measured and serve as a test of the approach and for parameter determination), and then to the decays of the ground-state axial-tensor mesons (that are mostly experimentally unknown). For the reader's convenience, in Table \ref{parameters} we summarize the parameters of the eLSM related to the (axial-)tensor sector that will be obtained via various fit procedures later on.

\begin{table}[h]
\begin{tabular}{|c|c|c|}
\hline
Parameters & Numerical Values & Fitting Data \\ \hline
   $h_3^{\text{ten}}$        &   $-41$               & Table \ref{masses}              \\ \hline
         $\delta_S^{\text{ten}}$  &   $0.3\,\text{GeV}^2$               & Table \ref{masses}             \\ \hline
       $g_{2}^{\text{ten}}$    & $(1.392\pm0.024)\cdot10^{4}(\text{MeV})$                 &        Table \ref{tabrtpp}      \\ \hline
         $g_{2}^{\prime\,\text{ten}}$  &    $(0.024\pm0.041)\cdot10^{4}(\text{MeV})$              &     Table \ref{tabrtpp}         \\ \hline
$g^{\text{ten}}_{2\,\text{lat}}$           &  $(0.7\pm 0.4)\cdot10^{4}(\text{MeV})$                &   Table \ref{tabkavpLattice}           \\ \hline
$c^{\text{ten}}$           &   $(4.8\pm0.9)\cdot10^{-7}(\text{MeV})^{-3}$               & Table \ref{tab:decay-tvp}             \\ \hline
$a^{\text{ten}}$           &   $(-2.09\pm 0.06)\cdot 10^{-2}(\text{MeV})^{-1}$               &  Table \ref{twogamma}            \\ \hline
    $a^{\prime\,\text{ten}}$       &   $(3.5\pm 0.4)\cdot 10^{-3}(\text{MeV})^{-1}$               & Table \ref{twogamma}             \\ \hline
    $\beta_T$       &  $(3.16\pm0.81)^\circ$                &        Table \ref{tabrtpp}      \\ \hline
\end{tabular}
\caption{Parameters of the eLSM connected to the (axial-)tensor mesons as determined by fits to known experimental data (see the following for details). Note, the suffix ``lat'' in $g^{\text{ten}}_{2\,\text{lat}}$ indicates that it was obtained via a comparison to lattice results. }
\label{parameters}
\end{table}

\subsection{Masses of (axial-)tensor mesons}
The expressions for the masses of tensor mesons are obtained by expanding Eq. (\ref{eq:mLag}) and taking into account the v.e.v.s and shifts of Eqs. (\ref{vev}) and (\ref{shiftedP}):
\begin{align}
    m_{\bold{a}_2}^2=\big( m^2_{\text{ten}}+2\delta_N^{\text{ten}}\big)+\frac{ h_3^{\text{ten}}\phi_N^2}{2}+\frac{h_2^{\text{ten}}\phi_N^2}{2}+\frac{h_1^{\text{ten}}}{2}(\phi_N^2+\phi_S^2)\,,
\end{align}
\begin{align}
    m_{K_{2}}^2=\big( m^2_{\text{ten}}+\delta_N^{\text{ten}}+\delta_S^{\text{ten}} \big) +\frac{h_1^{\text{ten}}}{2}(\phi_N^2+\phi_S^2)+\frac{1}{\sqrt{2}} h_3^{\text{ten}}\phi_N\phi_S+\frac{h_2^{\text{ten}}}{4}(\phi_N^2+2\phi_S^2)\,,
\end{align}
\begin{align}
    m_{f_{2,N}}^2=\big(m^2_{\text{ten}}+2\delta_N^{\text{ten}}\big)+\frac{ h_3^{\text{ten}}\phi_N^2}{2}+\frac{h_1^{\text{ten}}}{2}(\phi_N^2+\phi_S^2)+\frac{h_2^{\text{ten}}\phi_N^2}{2}\,,
\end{align}
\begin{align}
    m_{f_{2,S}}^2=\big( m^2_{\text{ten}}+2\delta_S^{\text{ten}}\big) +\frac{h_1^{\text{ten}}}{2}(\phi_N^2+\phi_S^2)+ h_3^{\text{ten}}\phi_S^2+h_2^{\text{ten}}\phi_S^2\,.
\end{align}
The analogous expressions for the axial-tensor mesons read:
\begin{align}
    m_{\rho_2}^2=\big( m^2_{\text{ten}}+2\delta_N^{\text{ten}}\big)-\frac{ h_3^{\text{ten}}\phi_N^2}{2}+\frac{h_2^{\text{ten}}\phi_N^2}{2}+\frac{h_1^{\text{ten}}}{2}(\phi_N^2+\phi_S^2)\,,
\end{align}
\begin{align}
    m_{K_{2A}}^2=\big( m^2_{\text{ten}}+\delta_N^{\text{ten}}+\delta_S^{\text{ten}} \big)+\frac{h_1^{\text{ten}}}{2}(\phi_N^2+\phi_S^2)-\frac{1}{\sqrt{2}} h_3^{\text{ten}}\phi_N\phi_S+\frac{h_2^{\text{ten}}}{4}(\phi_N^2+2\phi_S^2)\,,
\end{align}
\begin{align}
    m_{\omega_{2,N}}^2=\big( m^2_{\text{ten}}+2\delta_N^{\text{ten}}\big)-\frac{ h_3^{\text{ten}}\phi_N^2}{2}+\frac{h_2^{\text{ten}}\phi_N^2}{2}+\frac{h_1^{\text{ten}}}{2}(\phi_N^2+\phi_S^2)\,,
\end{align}
\begin{align}
    m_{\omega_{2,S}}^2=\big( m^2_{\text{ten}}+2\delta_S^{\text{ten}}\big)- h_3^{\text{ten}}\phi_S^2+h_2^{\text{ten}}\phi_S^2+\frac{h_1^{\text{ten}}}{2}(\phi_N^2+\phi_S^2)\,.
\end{align}
We observe that in each nonet the masses of the isovector and the nonstrange isoscalar members are identical:
\begin{equation}
    m_{\rho_2}^2= m_{\omega_{2,N}}^2\,\,\,\text{and}\,\,\, m_{\bold{a}_2}^2= m_{f_{2,N}}^2
    \text{ .}
\end{equation}
This degeneracy is broken by the (small) isoscalar mixing angle.

The following three equations relate the masses of chiral partners:
\begin{align}
m_{\rho_2}^2&= m_{\bold{a}_2}^2- h_3^{\text{ten}}\phi_N^2,\\
m_{K_{2A}}^2&= m_{K_{2}}^2-\sqrt{2} h_3^{\text{ten}}\phi_N\phi_S,\\
m_{\omega_{2,S}}^2&=m_{f_{2,S}}^2-2 h_3^{\text{ten}}\phi_S^2 \text{ ,}
\end{align}
out of which one obtains the following numerical estimate (using $K_2(1820)$ and $K_2^{*}(1430)$ as inputs): 
\begin{equation}\label{eq:h3}
    h_3^{\text{ten}}=\frac{m_{K_{2A}}^2-m_{K_{2}}^2}{\sqrt{2}\phi_N\phi_S} \simeq - 41\; \text{ .}
\end{equation}

Next, by using the PDG masses of the two tensor mesons and considering $\delta_N^{\text{ten}}\simeq 0$ and  $\phi_N^2\simeq 2\phi_S^2$, one gets:  
\begin{equation}
  \delta_S^{\text{ten}}= m_{K_{2}}^2-m_{\bold{a}_2}^2 \simeq 0.3\,\text{GeV}^2
  \text{ .}
\end{equation}
The following considerations are in order:
\begin{enumerate}
\item The mass of the purely strange resonance $\omega_{2,S} \simeq \phi_2$ is predicted to be:
\begin{equation}
   m_{\phi_{2}}^2 \simeq m_{\omega_{2,S}}^2= m_{\bold{a}_2}^2+2\delta_S^{\text{ten}}-\frac{3}{2} h_3^{\text{ten}}\phi_N^2\,,
\end{equation}
which implies the numerical value for the missing resonance $\phi_2 \simeq \omega_{2,S}$ as:
\begin{equation}
     m_{\phi_{2}}\approx 1971\,\text{MeV .} 
\end{equation}
In other words, the missing state $\phi_2(???)$ of Table 1 could appear as a novel resonance $\phi_2(1971)$.
    \item Using the following relation between the masses of chiral partners
    \begin{align*}
    m_{\rho_2}^2= m_{\bold{a}_2}^2- h_3^{\text{ten}}\phi_N^2
    \text{ ,}
\end{align*}
we get the following prediction for the not-yet discovered resonance $\rho_{2}$:
\begin{equation}
     m_{\rho_2}= 1663\,\text{MeV .} 
\end{equation}
 Since we neglect the isoscalar mixing, we also have $m_{\rho_2}\simeq 1663$ MeV. Thus, the new resonances $\rho_2(1663)$ and $\omega_2(1663)$ could possibly appear in the mesonic spectrum.
\item As a consistency check, the eLSM predicts the mass of the purely strange isoscalar tensor meson as
\begin{equation}
      m_{f_{2,S}}^2= m_{\omega_{2,S}}^2+2 h_3^{\text{ten}}\phi_S^2 \text{ ,}
\end{equation}
out of which $m_{f_{2,S}} \simeq 1520$ MeV,
slightly lighter than the physical value of the resonance $f_2^{\prime}(1525)$ that amounts to $1538\,\text{MeV}$. According to the relation $m_{\bold{a}_2}^2= m_{f_{2,N}}^2$, 
one has $m_{f_{2,N}}=1317\,\text{MeV}$, to be compared with the mass of $f_2(1270)$ of about $1297$ MeV. 

The small departures of the two isoscalar-tensor states can be easily explained by introducing a small isoscalar mixing angle. In our case, we shall study this mixing angle in the context of two-pseudoscalar and two-photon decay processes. 
When including the isoscalar mixing in the tensor sector, Table \ref{masses} summarizes our results for the (axial-)tensor masses. 
\end{enumerate}

\begin{table}[h]
\centering
\renewcommand{\arraystretch}{1.1}
\begin{tabular}
[c]{|l|c|c|c|}\hline
Resonances &  Masses (in \text{MeV})& Resonances &  Masses (in \text{MeV})\\\hline\hline
$\,\;a_2(1320)$ & \textbf{1317} & $\,\;\rho_2(?)$ & $1663$\\\hline
$\,\;K_2^\ast(1430)
$ &\textbf{1427} & $\,\;K_2(1820)
$ & \textbf{1819} \\\hline 
$\,\; f_2(1270)$ & $1297%
$ & $\,\; \omega_{2,N}$ & $1663%
$\\\hline
$\,\;f_2^\prime(1525)$ & $1538$ & $\,\;\omega_{2,S}$ & $1971$\\\hline
\end{tabular}
\caption{Masses of the mesons predicted by the eLSM. Bold entries are the PDG  values. The (small) mixing for the tensor-isoscalar mesons has been taken into account.}\label{masses}%
\end{table}
\subsection{Decays of tensor mesons}
\label{subsec:tensdec}
In this subsection, we compare the predictions for the strong and radiative decay of tensor mesons with available PDG data. 

\subsubsection{$T\longrightarrow P^{(1)}+P^{(2)}$ decay mode}
The interaction Lagrangian describing the decay of spin-$2$ tensor mesons to two pseudoscalar mesons follows from the chiral Lagrangian of Eq. (\ref{lag1}) by applying the shift of Eq. (\ref{shiftedP}):
\begin{equation}\label{g2ten}
\mathcal{L}_{tpp}=g_{2}^\text{ten}\,\mathrm{Tr}\Big[T^{\mu\nu}\big\{(\partial_\mu \mathcal{P})\,, (\partial_\nu \mathcal{P}) \big\}\Big]+\frac{g_{2}^{\prime\,\text{ten}}}{3}  \mathrm{Tr}\Big[T^{\mu\nu}\Big]\mathrm{Tr}\Big[\big\{(\partial_\mu \mathcal{P})\,, (\partial_\nu \mathcal{P}) \big\}\Big]\, \text{ .}
\end{equation}
The factor $1/3$ in front of the second term is chosen in such a way that for the singlet state the same normalization occurs. 
Namely, by using the decomposition $\sqrt{2}T_{2,\mu \nu }=\sum_{a=0}^{8}T_{\mu
\nu }^{a}\frac{\lambda ^{a}}{\sqrt{2}}$ together with $\lambda ^{0}=\sqrt{%
\frac{2}{3}}\mathbf{1,}$ it follows $T_{\mu \nu }=\frac{1}{\sqrt{6}}%
T_{\mu \nu }^{0}\mathbf{1+...}$ . For the singlet state the Lagrangian
reduces to $\mathcal{L}_{tpp}=(g_{2}^{\text{ten}}+g_{2}^{\prime \text{ten}%
})T^{0\,\mu \nu }\,\mathrm{Tr}\Big[\big\{(\partial _{\mu }%
\mathcal{P})\,,(\partial _{\nu }\mathcal{P})\big\}\Big]+...$ . This convention (which of course does not affect the results) is used in other terms as well. 

The tree-level decay rate is
\begin{equation}
\Gamma_{T\longrightarrow P^{(1)}+P^{(2)}}^{tl}(m_{t},m_{p^{(1)}},m_{p^{(2)}}%
)=\kappa_{tpp\,,i}\,\frac{ |\vec{k}_{p^{(1)},p^{(2)}}|^{5}}{60\,\pi\,m_{t}^{2}}%
\Theta(m_{t}-m_{p^{(1)}}-m_{p^{(2)}})\, \text{ ,} 
\label{eq:dec-tpp}%
\end{equation}
where the coefficients $\kappa_{tpp\,,i}$ (with dimension [Energy$^{-2}$]) are listed  in\ Table \ref{tabkpp}.
\begin{table}[h]
\centering
\renewcommand{\arraystretch}{1.5} 
\begin{tabular}{|l|c|}
\hline
Decay process   & $\kappa_{tpp\,,i}$ \\ \hline\hline
	$a_2(1320) \longrightarrow  \bar{K}\, K$ & $ 2 \Big(\frac{g_2^\text{ten}Z_k^2w_k^2}{2}\Big)^2$
\\ \hline
	$a_2(1320) \longrightarrow \pi\,\eta $  & $\Big( g_2^\text{ten}Z_\pi Z_{\eta_N}w_\pi w_{\eta_N}\cos\beta_P\Big)^2$
\\ \hline
$a_2(1320) \longrightarrow  \pi\,\eta^\prime  $ & $\Big(-g_2^\text{ten} Z_\pi Z_{\eta_N}w_\pi w_{\eta_N}\sin\beta_P\Big)^2$
\\ \hline
\hline
	$K_2^\ast(1430) \longrightarrow \pi\, \bar{K}$ & $ 3\Big(\frac{g_2^\text{ten}Z_kw_kZ_\pi w_\pi }{2}\Big)^2$ \\ \hline
\hline
	$f_2(1270) \longrightarrow \bar{K}\, K $ & $2\Big(\frac{Z_k^2w_k^2\big((4g_{2}^{\prime\,\text{ten}}+3g_2^\text{ten})\cos\beta_T+\sqrt{2}(2g_{2}^{\prime\,\text{ten}}+3g_2^\text{ten})\sin\beta_T\big)}{6}\Big)^2$
\\ 
\hline	$f_2(1270) \longrightarrow  \pi\,\pi$  & $6 \Big(\frac{Z_\pi^2w_\pi^2\big((2g_{2}^{\prime\,\text{ten}}+3g_2^\text{ten})\cos\beta_T+\sqrt{2}g_{2}^{\prime\,\text{ten}}\sin\beta_T\big)}{6}\Big)^2$
\\ \hline
$f_2(1270) \longrightarrow \eta \, \eta$  & $2 \Big(\frac{Z_{\eta_N}^2w_{\eta_N}^2\cos\beta_P^2 \big((2g_{2}^{\prime\,\text{ten}}+3g_2^\text{ten})\cos\beta_T+\sqrt{2}g_{2}^{\prime\,\text{ten}}\sin\beta_T\big)+Z_{\eta_S}^2w_{\eta_S}^2\sin\beta_P^2\big(\sqrt{2}(g_{2}^{\prime\,\text{ten}}+3g_2^\text{ten})\sin\beta_T+2g_{2}^{\prime\,\text{ten}}\cos\beta_T\big)}{6}\Big)^2$
\\ \hline
\hline
$f_2^\prime(1525) \longrightarrow \bar{K}\, K $  & $2\Big(\frac{Z_k^2w_k^2\big(-(4g_{2}^{\prime\,\text{ten}}+3g_2^\text{ten})\sin\beta_T+\sqrt{2}(2g_{2}^{\prime\,\text{ten}}+3g_2^\text{ten})\cos\beta_T\big)}{6}\Big)^2$
\\ \hline
	$f_2^\prime(1525) \longrightarrow  \pi\,\pi$    & $6 \Big(\frac{Z_\pi^2w_\pi^2\big(-(2g_{2}^{\prime\,\text{ten}}+3g_2^\text{ten})\sin\beta_T+\sqrt{2}g_{2}^{\prime\,\text{ten}}\cos\beta_T\big)}{6}\Big)^2$
\\ \hline
$f_2^\prime(1525) \longrightarrow \eta \, \eta$   & $2 \Big(\frac{Z_{\eta_N}^2w_{\eta_N}^2\cos\beta_P^2 \big(-(2g_{2}^{\prime\,\text{ten}}+3g_2^\text{ten})\sin\beta_T+\sqrt{2}g_{2}^{\prime\,\text{ten}}\cos\beta_T\big)+Z_{\eta_S}^2w_{\eta_S}^2\sin\beta_P^2\big(\sqrt{2}(g_{2}^{\prime\,\text{ten}}+3g_2^\text{ten})\cos\beta_T-2g_{2}^{\prime\,\text{ten}}\sin\beta_T\big)}{6}\Big)^2$
\\ \hline
\end{tabular}%
\caption{Coefficients for the decay channel $T\longrightarrow P^{(1)}+P^{(2)}$.}\label{tabkpp}
\end{table}

We perform a fit to the PDG data of Table \ref{tabrtpp} with three parameters: two coupling constants, the large-$N_c$ dominant $g_2^\text{ten}$ and the subdominant $g_2^{\prime\,\text{ten}}$, as well as the mixing anlge $\beta_T$. The fit gives:
\begin{equation}\label{g2}
g_2^\text{ten}=(1.392\pm0.024)\cdot10^{4}(\text{MeV})\,,\qquad g_2^{\prime\,\text{ten}}=(0.024\pm0.041)\cdot10^{4}(\text{MeV})\,,\qquad \beta_T= (3.16\pm0.81)^\circ \text{ .}
\end{equation}

\begin{table}[h]
\centering
\renewcommand{\arraystretch}{1.5} 
\begin{tabular}{|l|c|c|}
\hline
Decay process (in model)  & eLSM  (MeV)   &  PDG  (MeV)\\ \hline\hline
	$\,\;a_2(1320) \longrightarrow  \bar{K}\, K  $ &  $4.06\pm 0.14$  & $7.0^{+2.0}_{-1.5}\leftrightarrow (4.9 \pm 0.8) \%$
\\ \hline
	$\,\;a_2(1320) \longrightarrow \pi\,\eta   $  &  $25.37\pm 0.87$   & $18.5\pm 3.0\leftrightarrow (14.5 \pm 1.2) \%$ 
\\ \hline
$\,\;a_2(1320) \longrightarrow  \pi\,\eta^\prime (958)  $ &  $1.01\pm 0.03$   & $0.58\pm 0.10\leftrightarrow (0.55 \pm 0.09)\%$
\\ \hline
\hline
	$\,\;K_2^\ast(1430) \longrightarrow \pi\, \bar{K}   $ &  $44.82\pm 1.54$    & $49.9\pm 1.9\leftrightarrow (49.9 \pm 0.6)\%$\\ \hline
\hline
	$\,\;f_2(1270) \longrightarrow \bar{K}\, K $ &  $3.54\pm 0.29$    & $8.5\pm 0.8\leftrightarrow (4.6^{+0.5}_{-0.4})\%$ 
\\ \hline
	$\,\;f_2(1270) \longrightarrow  \pi\,\pi$  &  $168.82\pm 3.89$   &  $157.2^{+4.0}_{-1.1}\leftrightarrow (84.2^{+2.9}_{-0.9})\%$  
\\ \hline
$\,\;f_2(1270) \longrightarrow \eta \, \eta$  &  $0.67\pm0.03$ &    $0.75\pm 0.14 \leftrightarrow (0.4\pm 0.08)\%$ 
\\ \hline
\hline
$\,\;f_2^\prime(1525) \longrightarrow \bar{K}\, K   $  &  $23.72\pm 0.60$   & $75\pm 4\leftrightarrow (87.6 \pm 2.2)\%$
\\ \hline
	$\,\;f_2^\prime(1525) \longrightarrow  \pi\,\pi   $   &  $0.67\pm 0.14$  &  $0.71\pm 0.14\leftrightarrow (0.83 \pm 0.16)\%$
\\ \hline
$\,\;f_2^\prime(1525) \longrightarrow \eta \, \eta $   &  $1.81\pm 0.05$   & $9.9\pm 1.9 \leftrightarrow (11.6 \pm 2.2)\%$
\\ \hline
\end{tabular}%
\caption{Decay rates for $T\longrightarrow P^{
(1)} + P^{(2)}$.
}\label{tabrtpp}
\end{table}
As predicted by large-$N_c$ arguments, the results confirm that the coupling constant $g_2^{\prime\,\text{ten}}$ is much smaller than $g_2^{\text{ten}}$. Due to the quite large error, it is also compatible with zero. Yet, the phenomenology of large-$N_c$ suppressed decay (as e.g. the decays of the $J/\psi$ \cite{Kopke:1988cs}) shows that such terms, even if small, do not vanish.  
Moreover, the following prediction is obtained
\begin{equation}
   \text{Th}:\; \Gamma\Big[K_2^\ast(1430) \longrightarrow \eta\, \bar{K}\Big]=(1.13\pm 0.03)\cdot 10^{-3}\; \text{MeV},\;\;  \text{PDG}:\; \;0.15^{+0.34}_{-0.10}\; \text{MeV.}
\end{equation}
which due to large uncertainties was excluded from the fitting procedure.

Note, the isospin mixing angle $\beta_{T}=3.16^\circ$  is close (albeit not equal) to the value $\beta_{T}=5.7^\circ$ of PDG calculated via the following formula
\begin{align}
  \beta_{T}=35.3^\circ-\arctan{\Bigg(\sqrt{\frac{4m_{K_2}^2-m_{a_2}^2-3m_{f_2^\prime}^2}{-4m_{K_2}^2+m_{a_2}^2+3m_{f_2}^2}}\Bigg)}  \text{ .}
\end{align}

In general, the results of Table \ref{tabrtpp} show a good qualitative agreement. Yet, some entries are at odds with experimental ones, but one should recall that the experimental data are quite precise and only three parameters were used. The results could be improved by adding further large-$N_c$ suppressed terms as well as flavor-symmetry breaking contributions. In this way a more accurate description of the tensor decays could be possible. 
Such a precision study of tensor states could be interesting on its own but is not the main goal of this paper, thus it is left for the future.

\subsubsection{$T\longrightarrow V+P$ decay mode}

The next experimentally well-known decay channel is the one into a vector-pseudoscalar pair. Upon considering the shift of Eq. (\ref{vev}), the chiral Lagrangian of Eq. (\ref{chiralag-vp}) contains the term: 
\begin{align}\label{eq:lag-tvp}
    \mathcal{L}_{tvp}=(c_1^{\text{ten}}+c_2^{\text{ten}})\,\varepsilon_{\mu\nu\rho\sigma}\,\text{Tr}\Big[\,\partial^{\mu}T^{\nu\alpha}\,\Big(\partial^\rho V^{\sigma}(\partial_{\alpha}P)\Phi_0-\Phi_0(\partial_{\alpha}P\partial^\rho V^{\sigma})\Big)\Big] \text{ .}
\end{align}
 Defining $c^{\text{ten}}:=c_1^{\text{ten}}+c_2^{\text{ten}}$, the tree-level decay rate formula reads
\begin{equation}
\Gamma_{T\longrightarrow V+P}^{tl}(m_{t},m_{v},m_{p})=\frac{c^{\text{ten}\,2}\,|\vec{k}_{v,p}%
|^5}{40\,\pi}\,\kappa_{tvp\,,i}\,\Theta
(m_{t}-m_{v}-m_{p})\,,\label{eq:decaytvp}%
\end{equation}
where the factors $\kappa_{tvp\,,i}$ of mass dimension 2 are reported in Table \ref{tab:decay-tvp}.

 We obtain the value for the coupling $c^{\text{ten}}$ by performing a $\chi^2$-fit to the data of Table \ref{tab:decay-tvp} (for a brief overview of this procedure including the determination of the error of the obtained parameter(s), see the Appendix of Ref. \cite{Piotrowska:2017rgt}):
\begin{equation}
c^{\text{ten}}=(4.8\pm0.9)\cdot10^{-7}(\text{MeV})^{-3}%
\label{cten}
\text{ .}
\end{equation}
In the very same table the theoretical results are also presented. 
\begin{table}[h]
\centering
\renewcommand{\arraystretch}{1.3} 
\begin{tabular}{|l|c|c|c|}
\hline
Decay process (in model)  & eLSM (MeV) &  PDG-2020 (MeV) & $\kappa_{tvp\,,i}$\\ \hline\hline
	$\,\;a_2(1320) \longrightarrow  \rho(770)\,\pi $ &  $71.0\pm 2.6$ &  $73.61\pm 3.35\leftrightarrow (70.1 \pm 2.7) \%$ &  $2\Big(\frac{\phi_N}{4}\Big)^2$\\
\hline
$\,\;K_2^\ast(1430) \longrightarrow \bar{K}^{\ast}(892)\,\pi   $ &  $27.9\pm 1.0$  &  $26.92 \pm 2.14\leftrightarrow (24.7 \pm 1.6)\%$ &  $3\Big(\frac{\phi_N}{8}\Big)^2$ \\ \hline
	$\,\;K_2^\ast(1430) \longrightarrow \rho(770)\,K   $ &  $10.3\pm 0.4$  &  $9.48\pm 0.97\leftrightarrow (8.7 \pm 0.8)\%$ &  $3\Big(\frac{\sqrt{2}\phi_S}{8}\Big)^2$ \\ \hline
	$\,\;K_2^\ast(1430) \longrightarrow \omega(782)\,\bar{K}   $ &  $3.5\pm 0.1$  &  $3.16\pm 0.88\leftrightarrow (2.9 \pm 0.8)\%$  &  $\Big(\frac{-\phi_S\cos{\beta_V}+\phi_N\sin{\beta_V}}{4\sqrt{2}})\Big)^2$ \\ \hline
$\,\;f_2^\prime(1525) \longrightarrow \bar{K}^{\ast}(892)\,K+\mathrm{c}%
.\mathrm{c}. $ & $19.89\pm0.73$  &  & $4\Big(\frac{2\phi_S\cos{\beta_T}+\phi_N\sin{\beta_T}}{8})\Big)^2$   
\\ \hline
\end{tabular}%
\caption{ Decay rates and coefficients for the decay channel $T\longrightarrow P+V$. }\label{tab:decay-tvp}
\end{table}%
The model results are in good agreement with the experimental data. Moreover, a noteworthy prediction
concerning $f_2^\prime(1525) \rightarrow \bar{K}^{\ast}(892)\,K+\mathrm{c}%
.\mathrm{c}. $ is obtained.

\subsection{Radiative decays}

Radiative decays can be obtained via vector meson dominance \cite{OConnell:1995nse}
by applying the additional shift
\begin{equation}
V_{\mu\nu}\rightarrow V_{\mu\nu}+\frac{e}{g_{\rho}}F_{\mu\nu}Q\text{
,}\label{shiftvmd}%
\end{equation}
where $V_{\mu\nu}\equiv\partial_{\mu}V_{\nu}-\partial_{\nu}V_{\mu}$, $Q=\text{diag}\{2/3,-1/3,-1/3\}$ is the quark charge matrix, $F_{\mu\nu}=\partial_{\mu}A_{\nu}-\partial_{\nu}A_{\mu}$ is the
electromagnetic field tensor, $e=\sqrt{4\pi\alpha}$ is the electric coupling
constant, and $g_{\rho}\simeq5.5 \pm 0.5$ parametrizes the photon-vector-meson
transition. As a first application, we consider $T\longrightarrow
\gamma P$ by applying the VMD shift of Eq. (\ref{shiftvmd}) into Eq. (\ref{eq:lag-tvp}) that leads to the Lagrangian
\begin{align}\label{eq:lag-tgamp}
    \mathcal{L}_{t\gamma p}=\frac{e}{g_{\rho}}c^{\text{ten}}\,\text{Tr}\Big[\,\partial^{\mu}T^{\nu\alpha}\,\Big(F_{\mu\nu}Q(\partial_{\alpha}P)\Phi_0-\Phi_0(\partial_{\alpha}PF_{\mu\nu}Q\Big)\Big] \text{ ,}
\end{align}
and to the corresponding decay rate
\begin{equation}
\Gamma_{T\longrightarrow \gamma+P}^{tl}(m_{t},m_{p})=\frac{c^{\text{ten}\,2}\,|\vec{k}_{\gamma,p}%
|^5}{40\,\pi}\,\kappa_{t\gamma p\,,i}\,\Theta
(m_{t}-m_{p})\, .\label{eq:decaytgammap}%
\end{equation}

\begin{table}[h]
\centering
\renewcommand{\arraystretch}{1.3} 
\begin{tabular}{|l|c|c|c|}
\hline
Decay process (in model)  & eLSM &  PDG-2020 & $\kappa_{t\gamma p\,,i}$\\ \hline\hline
	$\,\;K_{2}^{\pm}(1430)\longrightarrow\gamma\,K^{\pm} $ &  $1.12\pm 0.47\,\,\text{MeV}$ &  $0.24\pm 0.05\,\,\text{MeV}$ &  $\frac{e^2}{g_{\rho}^2}\Big(\frac{\phi_N+2\sqrt{2}\phi_S}{12}\Big)^2$\\
\hline
$\,\;K_{2}^{0}(1430)\longrightarrow\gamma\,K^{0}$ &  $5.1\pm 2.2\,\,\text{keV}$  &  $<5.4\,\,\text{keV}$  &  $\frac{e^2}{g_{\rho}^2}\Big(\frac{\phi_N-\sqrt{2}\phi_S}{12}\Big)^2$\\ \hline
	$\,\;a_{2}^{\pm}(1320)\longrightarrow\gamma\,\pi^{\pm}   $ &  $1.01\pm 0.43\,\,\text{MeV}$  &  $0.31\pm 0.03\,\,\text{MeV}$  &  $\frac{e^2}{g_{\rho}^2}\Big(\frac{\phi_N}{4}\Big)^2$\\ \hline
\end{tabular}%
\caption{ Decay rates and coefficients for the decay channel $T\longrightarrow \gamma+P$.  }\label{tab:decay-tgammap}
\end{table}%

Taking into account
 $c^{\text{ten}}$ in Eq. (\ref{cten}) as well as  $g_{\rho}=5.5\pm 0.5$, we use the following equation to calculate the theoretical errors in Table \ref{tab:decay-tgammap}:
 \begin{equation}
     \delta \Gamma_{T\rightarrow\gamma P}(m_{t},m_{p})=\sqrt{\Big(\frac{\partial  \Gamma_{T\rightarrow\gamma P} (m_{t},m_{p})}{\partial c^{\text{ten}} }\delta c^{\text{ten}}\Big)^2+\Big(\frac{\partial  \Gamma_{T\rightarrow\gamma P} (m_{t},m_{p})}{\partial g_\rho }\delta g_\rho\Big)^2} \text{ .}
 \end{equation}

Our results for the photon-vector decays are reported in Table \ref{tab:decay-tgammap}: they are somewhat larger but, in consideration of the large errors, in qualitative agreement with the PDG results. 

Next, we describe the two-photon decays. We note that the chiral Lagrangian of Eq. (\ref{lag2}) delivers the following TVV interaction term: 
\begin{equation}\label{drvtlag}
    \mathcal{L}_{tvv}=a^{\text{ten}}\text{Tr}\Big[T_{\mu\nu}\{V^{\mu}_{\beta},V^{\nu\beta}\}\Big]+\frac{a^{\prime\,\text{ten}}}{3}\text{Tr}\Big[T_{\mu\nu}\Big]\text{Tr}\big[\{V^{\mu}_{\beta},V^{\nu\beta}\}\Big]\,.
\end{equation}
Then, through the VMD shift (\ref{shiftvmd}) for both vector fields, we get the Lagrangian
\begin{equation}
    \mathcal{L}_{t\gamma\gamma}=a^{\text{ten}}\frac{e^2}{g_{\rho}^2}\text{Tr}\Big[T_{\mu\nu}\{QF^{\mu}_{\beta}\,,QF^{\nu\beta}\}\Big]+\frac{a^{\prime\,\text{ten}}}{3}\frac{e^2}{g_{\rho}^2}\text{Tr}\Big[T_{\mu\nu}\Big]\text{Tr}\Big[\{QF^{\mu}_{\beta}\,,QF^{\nu\beta}\}\Big]\text{ ,}
\end{equation}
and the decay rate of the type
\begin{equation}\label{eq:gamgam-decay}
\Gamma_{T\rightarrow \gamma\gamma}^{tl}=\frac{e^4}{g_{\rho}^4}\frac{2\,|\vec{k}%
|^5}{5\,\pi\,m_{t}^{2}}\kappa_{t\gamma\gamma\,,i}=\frac{e^4}{g_{\rho}^4}\frac{m_{t}^3}{80\pi}\kappa_{t\gamma\gamma\,,i}
\text{ ,}
\end{equation}
where $m_{t}=2|\vec{k}|$ was used. 

By using the experimental values of Table \ref{twogamma}, a standard $\chi^2$ approach with two free parameters (the coupling constants $a^{\text{ten}}$ and $a^{\prime\, \text{ten}}$, $\beta_T=3.16^\circ$ kept fixed) and three experimental data-points, gives: 
\begin{equation}
    a^{\text{ten}}=(-2.09\pm 0.06)\cdot 10^{-2}(\text{MeV})^{-1}\,,\qquad  a^{\prime\,\text{ten}}=(3.5\pm 0.4)\cdot 10^{-3}(\text{MeV})^{-1} 
    \label{agamma}
    \text{ }
\end{equation}
The theoretical results are also listed in Table \ref{twogamma}. 
 Thus, in this case the coupling constants are directly fitted to the decay rates. The fact that all 3 data points are well described confirm that the value the mixing angle $\beta_T=3.16^\circ$ is consistent with data. Moreover,
the parameter $a^{\prime\,\text{ten}}$ turns out to be much smaller than $a^{\text{ten}}$ in agreement with the large-$N_c$ expectations.
[Note: in the fit, we fixed $g_{\rho} = 5.5$ and determined the coupling constant $a^{\text{ten}}$; actually, the combination $a^{\text{ten}}/g_{\rho}^{2}$ enters into the expressions. 
Yet, allowing for a variation of $g_{\rho}$ would not lead to any significant changes here, since it could be reabsorbed by a change of $a^{\text{ten}}$.]

These radiative decay channels also show that the coupling with photons is sizable, thus the production of these not-yet discovered states $\rho_2$, $\omega_2$, and $\phi_2$ in photoproduction experiments is well upheld.

\begin{table}[h]
\centering
\renewcommand{\arraystretch}{1.3} 
\begin{tabular}{|l|c|c|c|}
\hline
Decay process (in model)  & eLSM (keV) &  PDG-2020 (keV)  & $\kappa_{t\gamma\gamma\,,i}$ \\ \hline\hline
	$\,\;a_2(1320) \longrightarrow  \gamma\gamma $ &  $1.01\pm 0.06$ &  $1.00\pm 0.06$ &  $\frac{a^{\text{ten}\,2}}{36}$\\
 \hline
 $\,\;f_2(1270) \longrightarrow \gamma\gamma $ & $1.95\pm 0.10$  & $2.6\pm0.5$ &  $\Big(\frac{(5 a^{\text{ten}} + 4 a^{\prime\,\text{ten}}) \cos{\beta_{T}} + \sqrt{2} (a^{\text{ten}} + 2 a^{\prime\,\text{ten}}) \sin{\beta_{T}} }{18}\Big)^2$
\\ \hline
$\,\;f_2^\prime(1525) \longrightarrow \gamma\gamma $ & $0.083\pm0.009$  & $0.082\pm0.009$ &  $\Big(\frac{-(5 a^{\text{ten}} + 4 a^{\prime\,\text{ten}}) \sin{\beta_{T}} + \sqrt{2} (a^{\text{ten}} + 2 a^{\prime\,\text{ten}}) \cos{\beta_{T}} }{18}\Big)^2$
\\ \hline
\end{tabular}%
\caption{Decay rates and coefficients for the decay channel $T\longrightarrow \gamma\gamma$ .  }
\label{twogamma}
\end{table}%

It is possible to use the two parameters presented in Eq. (\ref{agamma}) for the predictions of numerous photon-vector decay mode by shifting only one vector field of the Lagrangian of Equation \eqref{drvtlag}
\begin{equation}
    \mathcal{L}_{t\gamma V}=2a^{\text{ten}}\frac{e}{g_{\rho}}\text{Tr}\Big[T_{\mu\nu}\{QF^{\mu}_{\beta}\,,V^{\nu\beta}\}\Big]+\frac{2a^{\prime\,\text{ten}}}{3}\frac{e}{g_{\rho}}\text{Tr}\Big[T_{\mu\nu}\Big]\text{Tr}\Big[\{QF^{\mu}_{\beta}\,,V^{\nu\beta}\}\Big] \text{ .}\label{a2lag}
\end{equation}

The vector-photon decay rates take the form
\begin{align}\nonumber
 \Gamma_{T\longrightarrow \gamma+V}^{tl}(m_{t},m_{v})= \frac{\kappa_{t\gamma v\,,i}\,e^2}{15\,\pi\,m_{t}^{2}m_{v}^2g_\rho^2}\Big(3|\vec{k}_{v,\gamma}|^7&+5|\vec{k}_{v,\gamma}|^3m_v^2+5|\vec{k}_{v,\gamma}|^4m_v^2\sqrt{|\vec{k}_{v,\gamma}|^2+m_v^2}\\
 &\qquad+|\vec{k}_{v,\gamma}|^5\big(10m_v^2-3(|\vec{k}_{v,\gamma}|^2+m_v^2)\big)\Big)\, \text{ .}
\end{align}
The corresponding  coefficients are reported in Table \ref{tabrvgam} and the numerical predictions for the not-yet measured decay rates in Table \ref{tabrvgam2}. The corresponding errors are obtained by considering Eq. (\ref{agamma}) together with $g_{\rho} = 5.5 \pm 0.5$.
[Note, the variation for $g_{\rho} = 5.5 \pm 0.5$ must be taken into account for the evaluation of the errors since the different ratio $a^{\text{ten}}/g_{\rho}$ enters into the expressions.]
Additional sources of errors, such as the uncertainties of masses, were neglected for simplicity. Thus, the actual errors can be somewhat larger but of the same order of the ones reported in Table \ref{tabrvgam2}.
\begin{table}[h]
\centering
\renewcommand{\arraystretch}{1.1} 
\begin{tabular}{|l|c|}
\hline

Decay process (in model)  & $\kappa_{t\gamma v\,,i}$ \\ \hline\hline
	$\,\;a_2(1320) \longrightarrow  \rho(770)\,\gamma $ &  $\Big(\frac{a^{\text{ten}}}{6}\Big)^2$  \\
\hline
	$\,\;a_2^{0}(1320) \longrightarrow  \omega(782)\,\gamma $ &  $\Big(\frac{a^{\text{ten}}\cos{\beta_V}}{2}\Big)^2$  \\
\hline
	$\,\;a_2(1320) \longrightarrow  \phi(1020)\,\gamma $ &  $\Big(\frac{a^{\text{ten}}\sin{\beta_V}}{2}\Big)^2$  \\
\hline\hline
$\,\;K_2^{\ast\pm}(1430) \longrightarrow \bar{K}^{\ast\pm}(892)\,\gamma   $ &  $\Big(\frac{a^{\text{ten}}}{6}\Big)^2$   \\ \hline
$\,\;K_2^{\ast\; 0}(1430) \longrightarrow \bar{K}^{\ast\;0}(892)\,\gamma   $ &  $\Big(\frac{a^{\text{ten}}}{3}\Big)^2$
 \\ \hline\hline
$\,\;f_2(1270) \longrightarrow \rho(770)\,\gamma $ & $\Big(\frac{(a^{\text{ten}}+2a^{\prime\,\text{ten}})\cos{\beta_T}+\sqrt{2}a^{\prime\,\text{ten}}\sin{\beta_T}}{2}\Big)^2$   
\\ \hline
$\,\;f_2(1270) \longrightarrow \omega(782)\,\gamma $ & $\Big(\frac{\cos{\beta_V}\big((a^{\text{ten}}+2a^{\prime\,\text{ten}})\cos{\beta_T}+\sqrt{2}a^{\prime\,\text{ten}}\sin{\beta_T}\big)-2\sin{\beta_V}\big((a^{\text{ten}}+a^{\prime\,\text{ten}})\sin{\beta_T}+\sqrt{2}a^{\prime\,\text{ten}}\cos{\beta_T}\big)}{6}\Big)^2$   
\\ \hline
$\,\;f_2(1270) \longrightarrow \phi(1020)\,\gamma $ & $\Big(\frac{\sin{\beta_V}\big((a^{\text{ten}}+2a^{\prime\,\text{ten}})\cos{\beta_T}+\sqrt{2}a^{\prime\,\text{ten}}\sin{\beta_T}\big)+2\cos{\beta_V}\big((a^{\text{ten}}+a^{\prime\,\text{ten}})\sin{\beta_T}+\sqrt{2}a^{\prime\,\text{ten}}\cos{\beta_T}\big)}{6}\Big)^2$   
\\ \hline\hline
$\,\;f_2^\prime(1525) \longrightarrow \rho(770)\,\gamma $ & $\Big(\frac{-(a^{\text{ten}}+2a^{\prime\,\text{ten}})\sin{\beta_T}+\sqrt{2}a^{\prime\,\text{ten}}\cos{\beta_T}}{2}\Big)^2$   
\\ \hline
$\,\;f_2^\prime(1525) \longrightarrow \omega(782)\,\gamma $ & $\Big(\frac{\sin{\beta_T}\big(-(a^{\text{ten}} + 2a^{\prime\,\text{ten}}) \cos{\beta_V} + 
    2 \sqrt{2} a^{\prime\,\text{ten}} \sin{\beta_V}\big) + 
 \cos{\beta_T} \big(\sqrt{2} a^{\prime\,\text{ten}}\cos{\beta_V} - 2 (a^{\text{ten}} + a^{\prime\,\text{ten}}) \sin{\beta_V}\big)}{6}\Big)^2$   
\\ \hline
$\,\;f_2^\prime(1525) \longrightarrow \phi(1020)\,\gamma $ & $\Big(\frac{\cos{\beta_T} \big(-2 (a^{\text{ten}} + a^{\prime\,\text{ten}})\cos{\beta_V} -\sqrt{2} a^{\prime\,\text{ten}} \sin{\beta_V} \big)+ \sin{\beta_T} \big(2 \sqrt{2} a^{\prime\,\text{ten}} \cos{\beta_V} + (a^{\text{ten}} + 2 a^{\prime\,\text{ten}})\sin{\beta_V}\big)}{6}\Big)^2$   
\\ \hline
\end{tabular}%
\caption{ Coefficients for the decay channel $T\longrightarrow \gamma+V$ decays. }\label{tabrvgam}
\end{table}%

\begin{table}[h]
\centering
\renewcommand{\arraystretch}{1.1} 
\begin{tabular}{|l|c|}
\hline
Decay process (in model)  & Decay Width (MeV) \\ \hline\hline
	$\,\;a_2(1320) \longrightarrow  \rho(770)\,\gamma $ &  $0.22\pm 0.04$  \\
\hline
	$\,\;a_2^{0}(1320) \longrightarrow  \omega(782)\,\gamma $ &  $1.94\pm 0.04$  \\
\hline
	$\,\;a_2(1320) \longrightarrow  \phi(1020)\,\gamma $ &  $0.0024\pm 0.0005$  \\
\hline\hline
$\,\;K_2^{\ast\pm}(1430) \longrightarrow \bar{K}^{\ast\pm}(892)\,\gamma   $ &  $0.23\pm 0.04$   \\ \hline
$\,\;K_2^{\ast\; 0}(1430) \longrightarrow \bar{K}^{\ast\;0}(892)\,\gamma   $ &  $0.94\pm 0.18$
 \\ \hline\hline
$\,\;f_2(1270) \longrightarrow \rho(770)\,\gamma $ & $0.70\pm 0.17$   
\\ \hline
$\,\;f_2(1270) \longrightarrow \omega(782)\,\gamma $ & $0.068\pm 0.017$   
\\ \hline
$\,\;f_2(1270) \longrightarrow \phi(1020)\,\gamma $ & $0.007\pm 0.002$   
\\ \hline\hline
$\,\;f_2^\prime(1525) \longrightarrow \rho(770)\,\gamma $ & $0.32\pm 0.08$   
\\ \hline
$\,\;f_2^\prime(1525) \longrightarrow \omega(782)\,\gamma $ & $0.012\pm 0.005$   
\\ \hline
$\,\;f_2^\prime(1525) \longrightarrow \phi(1020)\,\gamma $ & $0.61\pm 0.12$   
\\ \hline
\end{tabular}%
\caption{ Predictions for $T\longrightarrow \gamma+V$ .}\label{tabrvgam2}
\end{table}%

\subsubsection{Vector-vector decay mode of tensor mesons}
We conclude the present subsection by considering the following PDG data:
\begin{align*}
    \text{PDG}:&\,\,\,\Gamma[f_2(1270)\longrightarrow \pi^{+}\pi^{-}2\pi^{0}] \approx 19.5 \,\text{MeV;}\\
    \text{PDG}:&\,\,\,\Gamma[a_2(1320)\longrightarrow \omega(782)\pi\pi] \approx 11.3 \,\text{MeV;}\\
    \text{PDG}:&\,\,\,\Gamma[K_2^\star(1430)\longrightarrow K^\star(892) \pi \pi] = 13.5\pm 2.3 \,\text{MeV.}
\end{align*}

It is natural to interpret these decays as vector-vector decay modes ($\rho(770) \rho(770)$, $\omega(782) \rho(770)$, and $K^{*}(892) \rho(770)$), in which the $\rho(770)$ meson further decays into two pions. 
In order to describe them, we introduce the properly normalized ``Sill'' spectral function discussed in Ref. \cite{Giacosa:2021mbz} for the $\rho$-meson as
\begin{align}
    d_{\rho}(y)=\frac{2y}{\pi}\frac{\sqrt{y^2-4m_{\pi}^2}\,\tilde{\Gamma}_{\rho}}{(y^2-m_{\rho}^2)^2+(\sqrt{y^2-4m_{\pi}^2}\,\tilde{\Gamma}_{\rho})^2}\,\Theta
(x-2m_{\pi}),\,\, \int_{0}^{\infty}dy\,d_{\rho}(y)=1
\text{ ,}
\end{align}
where $\tilde{\Gamma}_{\rho}$ is defined as 
\begin{align}
    \tilde{\Gamma}_{\rho}\equiv \frac{\Gamma_{\rho\rightarrow 2\pi}\,m_{\rho}}{\sqrt{m_{\rho}^2-4m_{\pi}^2}}\;
    \text{ ,}
\end{align}
in which $\Gamma_{\rho\rightarrow 2\pi} = 149.1$ MeV and $m_{\rho} =775.11 $ MeV are taken from the PDG (small errors omitted).

The Sill spectral function is useful because the normalization is guaranteed also for broad states and because the contribution of virtual particles (the pion-pion loop for the $\rho$ meson) has a vanishing real part, thus only the imaginary part describing the decay width is sufficient (in other words, no modification of the $\rho(770)$-mass part of the denominator appears), see details in Ref. \cite{Giacosa:2021mbz}. 
 
There are two chiral invariant Lagrangians, Eq. (\ref{lag1}) and Eq. (\ref{lag2}), that contribute to the coupling of tensor mesons to two vector mesons. We present the following approximate results by considering destructive interference between the corresponding amplitudes (the constructive case delivers too large decay rates): 
\begin{align}\label{decay}
    \Gamma_{f_2\rightarrow \rho\rho\rightarrow4\pi}\simeq \int_{0}^{\infty}dy_1\,d_{\rho}(y_1) \,\int_{0}^{\infty}dy_2\,d_{\rho}(y_2)\,\Gamma_{f_2\rightarrow \rho\rho}(m_{f_2},y_1,y_2)\approx29.9 \,\text{MeV ;}
\end{align}
\begin{align}\label{decay}
    \Gamma_{a_2\rightarrow \omega\rho\rightarrow\omega 2\pi}\simeq \int_{0}^{\infty}dy_1\,d_{\rho}(y_1) \,\Gamma_{a_2\rightarrow \rho\omega}(m_{\bold{a}_2},m_{\omega_1},y_1)\approx 11.1 \,\text{MeV ;}
\end{align}
\begin{align}\label{decay}
    \Gamma_{K_2\rightarrow K^\star\rho\rightarrow K 2\pi}\simeq \int_{0}^{\infty}dy_1\,d_{\rho}(y_1) \,\Gamma_{K_2\rightarrow \rho\omega}(m_{K_2},m_{K_1},y_1)\approx 6.6 \,\text{MeV .}
\end{align}
The results are in qualitative agreement with the experimental ones: this is quite remarkable since no additional parameter is requested. Moreover, this is a nontrivial process that involves two distinct interaction terms (with destructive interference) as well as the integral over the spectral function of the $\rho$-meson.


\subsection{Decays of axial-tensor mesons}
The ground-state mesons with $J^{PC}=2^{--}$ are elusive. In this section we use the eLSM to predict their decay rates and compare with the available lattice QCD (LQCD) and other theoretical results. 

\subsubsection{$A_2\longrightarrow V+P$ decay mode}
The chiral Lagrangian (\ref{lag1}) contains the decay of axial-vector states into a vector and
pseudoscalar pair: 
\begin{equation}
\mathcal{L}_{a_2vp}=g_{2}^{\text{ten}}\,\mathrm{Tr}\Big[A_2^{\mu\nu}\big\{(\partial_\mu \mathcal{P})\,, V_{\nu}\big\}\Big]\,\text{ .}%
\label{eq:lag-wsp}%
\end{equation}
The tree-level decay rate takes the form:
\begin{equation}
\Gamma_{A_2\longrightarrow V+P}^{tl}(m_{a_2},m_{v},m_{p})=\frac
{g_2^{\text{ten}\,2}\,|\vec{k}_{v,p}|^{3}}{120\,\pi\,m_{a_2}^{2}}
\Big(5+\frac{2\,|\vec{k}_{v,p}|^{2}}{m_{v}^{2}}\Big)\,\kappa
_{a_2vp\,,i}\,\Theta(m_{a_2}-m_{v}-m_{p})\,,
\end{equation}
where the $\kappa
_{a_2vp\,,i}$ can be found in Table \ref{tabkavp} together with the decay widths. We observe quite large decay rates for this decay mode. Yet, our results can be considered qualitative for various reasons: 

(i) the interaction strength is determined solely by the knowledge on tensor mesons: while this is in agreement with chiral symmetry, no corrections to it have been taken into account, but chiral symmetry breaking effects are expected to be relevant between 1-2 GeV;

(ii) the result depends on quantities of the type $(Zw)^{2} \propto g_1^{4}$ (see Tables \ref{tabkavp} and \ref{constants}), thus an even small uncertainty in this parameter of the original eLSM generates a quite large change of the results of the decays of axial-tensor states;

(iii) the parameter $g_1$ (as well as others, such as $g_2^{\text{ten}}$) is taken as a constant over a large range of energies, yet it is reasonable to assume that it has a (soft) energy dependence. One might expect that it becomes smaller when the mass of the decaying particle increases, thus embedding the effect of form factors (and eventually corrections due to quantum fluctuations). Also in this case, even a small change would have a non-negligible consequence leading to the left side of the quoted ranges in table \ref{tabkavp}.


Summarizing, the used parameters (in particular $g_1$) imply a quite large and difficult to determine source of uncertainty of our decays of axial-tensor states. In order to be on the safe side, we thus estimate a quite large error of about 50\%, for the reported numbers in Table \ref{tabkavp}. 

Experimental data is missing for the $\rho_2$, $\omega_{2,N} \simeq \omega_{2}$, and $\omega_{2,S} \simeq \phi_{2}$. In the kaonic sectors, the resonances $K_2(1820)$ and  $K_2(1770)$ are expected to emerge from the bare axial-tensor and pseudotensor mesons which can mix in a way that resembles the mixing of the axial-vector and pseudovector mesons described in Eq. (\ref{mixk1abfields}).  

According to the assignment of Table \ref{mesonlist}, $K_{2A}$ is closer to $K_2(1820)$, (but this is not yet settled).  The decay width for the $K_{2A}$ state described in our work is approximately twice of the experimentally measured decay of $K_2(1820)$. 
This fact suggests that the left sides of the range of Table \ref{tabkavp} for the $K_{2A}$ state is favoured. 

This interpretation can be extended to the other decay rates by a comparison with the lattice QCD results reported in Table \ref{tabkavpLattice}. Namely, our theoretical results for decays are of the same order of magnitude but about 2-3 times larger than the LQCD ones. Even though both approaches have large uncertainties, this comparison (as well as the previous arguments) signalizes that the low side of the range of each decay mode of the eLSM should be preferred.


For completeness we also perform a parameter determination by directly using the LQCD results, that leads to  $g^{\text{ten}}_{2\,\text{lat}} \simeq (0.7\pm 0.4)\cdot10^{4}(\text{MeV})$ (assuming 50 percent error in \cite{Johnson:2020ilc})\footnote{Note that,  in the $SU(3)$ flavor symmetry based LQCD calculation, 
the  lightest pseudoscalar states have a  mass of about 700 MeV. }. This value is within two sigma of the previous determination in Eq. (\ref{g2}).
In Table \ref{tabkavpLattice}, we present the corresponding results. Quite interestingly, the various entries are consistent with each other, thus showing that any ratio of decays constructed with lattice results is in agreement with the corresponding eLSM one. 

In summary, the outcome of our theoretical study is pretty clear: although the uncertainties are still large, the axial-tensor states are expected to be broad even when taking the low side of our results. This fact could eventually explain the missing experimental observation of the putative states $\rho_2$, $\omega_2$, and $\phi_2$ up to now.

 \begin{table}[h] 
		\centering
		\renewcommand{\arraystretch}{1.1}
		\begin{tabular}[c]{|l|c|c|}
			\hline
			Decay process (in model) & eLSM (MeV) &$\kappa_{a_2vp\,,i}$  \\
			\hline \hline
			$\,\;\rho_2(?) \longrightarrow \rho(770)\, \eta$ & $ \approx 99\pm 50$ & $\Big(- w_{\eta_N} Z_{\eta_N} \cos{\beta_P}\Big)^2$   \\
			\hline
			$\,\;\rho_2(?) \longrightarrow \bar{K}^\ast(892)\, K + \mathrm{c}.\mathrm{c}.$ & $  \approx 85\pm 43$ & $4\Big(\frac{Z_Kw_K}{2}\Big)^2$  \\
			\hline
			$\,\;\rho_2(?) \longrightarrow \omega(782)\, \pi$ & $ \approx 419\pm 210 $ & $\Big(- w_{\pi} Z_{\pi} \cos{\beta_V}\Big)^2$ \\
			\hline
			$\,\;\rho_2(?) \longrightarrow \phi(1020)\, \pi$ & $ \approx 0.8 $ & $\Big( w_{\pi} Z_{\pi} \sin{\beta_V}\Big)^2$ \\
			\hline \hline
			$\,\;K_{2,A} \longrightarrow \rho(770)\, K$ & $ \approx 195\pm 98$ & $3\Big(\frac{Z_Kw_K}{2}\Big)^2$ \\
			\hline
			$\,\;K_{2,A} \longrightarrow \bar{K}^\ast(892)\, \pi $ & $ \approx 316\pm 158 $ & $3\Big(\frac{Z_{\pi}w_{\pi}}{2}\Big)^2$ \\
			\hline
			$\,\;K_{2,A} \longrightarrow \bar{K}^\ast(892)\, \eta$ & $ \approx0.01$ & $\Big( -\frac{1}{2} (\sqrt{2}Z_{\eta_S}w_{\eta_S}\sin{\beta_P}+Z_{\eta_N}w_{\eta_N}\cos{\beta_P})\Big)^2$ \\
			\hline
			$\,\;K_{2,A} \longrightarrow \omega(782)\, \bar{K}$ & $ \approx 51\pm 26$ & $\Big( -\frac{w_{K} Z_{K}}{2} (\sqrt{2}\sin{\beta_V}+\cos{\beta_V})\Big)^2$ \\
			\hline
			$\,\;K_{2,A} \longrightarrow \phi(1020)\, \bar{K}$ & $  \approx 50\pm 25$ & $\Big( \frac{w_{K} Z_{K}}{2} (-\sqrt{2}\cos{\beta_V}+\sin{\beta_V})\Big)^2$ \\
			\hline \hline
			$\,\;\omega_{2,N} \longrightarrow \rho(770)\, \pi$ & $ \approx 1314\pm 657 $  & $3\Big(-Z_{\pi}w_{\pi}\Big)^2$ \\
			\hline
			$\,\;\omega_{2,N} \longrightarrow \bar{K}^\ast(892)\, K + \mathrm{c}.\mathrm{c}.$ & $ \approx 85\pm 43 $ & $4\Big(-\frac{Z_{K}w_{K}}{2}\Big)^2$  \\
			\hline
			$\,\;\omega_{2,N} \longrightarrow \omega(782)\, \eta$ & $ \approx 93\pm 47$ & $\Big( Z_{\eta_N}w_{\eta_N}\cos{\beta_P}\cos{\beta_V}\Big)^2$ \\
			\hline
			$\,\;\omega_{2,N} \longrightarrow \phi(1020)\, \eta$ & $ \approx 0.06 $  & $\Big( Z_{\eta_N}w_{\eta_N}\cos{\beta_P}\sin{\beta_V}\Big)^2$\\
				\hline \hline
					$\,\;\omega_{2,S} \longrightarrow \bar{K}^\ast(892)\, K + \mathrm{c}.\mathrm{c}.$ & $ \approx 510\pm 255$  & $4\Big(-\frac{Z_{K}w_{K}}{\sqrt{2}}\Big)^2$  \\
			\hline
			$\,\;\omega_{2,S} \longrightarrow \omega(782)\, \eta$ & $ \approx 1.0\pm0.5 $ & $\Big(-\sqrt{2} Z_{\eta_S}w_{\eta_S}\sin{\beta_P}\sin{\beta_V}\Big)^2$ \\
			\hline
			$\,\;\omega_{2,S} \longrightarrow \omega(782)\, \eta^\prime(958)$ & $ \approx 0.3 $ & $\Big(-\sqrt{2} Z_{\eta_S}w_{\eta_S}\cos{\beta_P}\sin{\beta_V}\Big)^2$ \\
			\hline
			$\,\;\omega_{2,S} \longrightarrow \phi(1020)\, \eta$ & $ \approx 101\pm 51 $ & $\Big(-\sqrt{2} Z_{\eta_S}w_{\eta_S}\cos{\beta_V}\sin{\beta_P}\Big)^2$  \\
			\hline
		\end{tabular}
		\caption{Predictions and coefficients for the decay channel $A_{2}\longrightarrow  V+P$.}\label{tabkavp}
		\end{table}

\begin{table}[h] 
		\centering
		\renewcommand{\arraystretch}{1.1}
		\begin{tabular}[c]{|l|c|c|}
			\hline
			Decay process (in model) & eLSM  (MeV) & LQCD  (MeV) \cite{Johnson:2020ilc}  \\
			\hline \hline
			$\,\;\rho_2(?) \longrightarrow \rho(770)\, \eta$ & $ \approx 30$ & $ $   \\
			\hline
			$\,\;\rho_2(?) \longrightarrow \bar{K}^\ast(892)\, K + \mathrm{c}.\mathrm{c}.$ & $  \approx 27$ & $36$  \\
			\hline
			$\,\;\rho_2(?) \longrightarrow \omega(782)\, \pi$ & $ \approx 122 $ & $125$ \\
			\hline
			$\,\;\rho_2(?) \longrightarrow \phi(1020)\, \pi$ & $ \approx 0.3 $ & $ $ \\
			\hline \hline
			$\,\;K_{2,A} \longrightarrow \rho(770)\, K$ & $ \approx 53$ & $ $ \\
			\hline
			$\,\;K_{2,A} \longrightarrow \bar{K}^\ast(892)\, \pi $ & $ \approx 87 $ & $ $ \\
			\hline
			$\,\;K_{2,A} \longrightarrow \bar{K}^\ast(892)\, \eta$ & $ \approx0.004$ & $ $ \\
			\hline
			$\,\;K_{2,A} \longrightarrow \omega(782)\, \bar{K}$ & $ \approx 13.8$ & $ $ \\
			\hline
			$\,\;K_{2,A} \longrightarrow \phi(1020)\, \bar{K}$ & $  \approx 13.7$ & $ $ \\
			\hline \hline
			$\,\;\omega_{2,N} \longrightarrow \rho(770)\, \pi$ & $ \approx 363 $  & $365$ \\
			\hline
			$\,\;\omega_{2,N} \longrightarrow \bar{K}^\ast(892)\, K + \mathrm{c}.\mathrm{c}.$ & $ \approx 25 $ & $36$  \\
			\hline
			$\,\;\omega_{2,N} \longrightarrow \omega(782)\, \eta$ & $ \approx 27$ & $17$ \\
			\hline
			$\,\;\omega_{2,N} \longrightarrow \phi(1020)\, \eta$ & $\approx 0.02 $ & $  $\\
				\hline \hline
					$\,\;\omega_{2,S} \longrightarrow \bar{K}^\ast(892)\, K + \mathrm{c}.\mathrm{c}.$ & $ \approx 100$  & $148$  \\
			\hline
			$\,\;\omega_{2,S} \longrightarrow \omega(782)\, \eta$ & $ \approx 0.2 $ & $ $ \\
			\hline
			$\,\;\omega_{2,S} \longrightarrow \omega(782)\, \eta^\prime(958)$ & $ \approx 0.02 $ & $ $ \\
			\hline
			$\,\;\omega_{2,S} \longrightarrow \phi(1020)\, \eta$ & $ \approx 17 $ & $44$  \\
			\hline
		\end{tabular}
		\caption{ Predictions for the $A_{2}\longrightarrow  V+P$ based on LQCD results. }\label{tabkavpLattice}
		\end{table}

\subsubsection{$ A_2\longrightarrow A_{1}+P$ decay mode}

The chiral Lagrangian of Eq. (\ref{chiralag-vp}) contains the decay into an axial-tensor and pseudoscalar pair:
\begin{align}
    \mathcal{L}_{a_2a_1p}=c^{\prime\,\text{ten}}\,\varepsilon_{\mu\nu\rho\sigma}\,\text{Tr}\Big[\,\partial^{\mu}A_2^{\nu\alpha}\,\Big((\partial^\rho A_1^{\sigma})(\partial_{\alpha}P)\Phi_0-\Phi_0(\partial_{\alpha}P)(\partial^\rho A_1^{\sigma})\Big)\Big] \text{ ,}
\end{align}
where $c^{\prime\,\text{ten}}:=c_1^{\text{ten}}-c_2^{\text{ten}}$. The corresponding decay is: 
\begin{equation}
\Gamma_{A_2\longrightarrow A_1+P}^{tl}(m_{a_2},m_{a_1},m_{p})=\frac{c^{\prime\,\text{ten}\,2}\,|\vec{k}_{a_1,p}%
|^5}{40\,\pi}\,\kappa_{a_2ap\,,i}\Theta
(m_{a_2}-m_{a_1}-m_{p})\, \text{ .}
\label{eq:decay}%
\end{equation}
The theoretical predictions (as well as the $\kappa_{a_2ap\,,i}$ values)  for each channel proportional to the factor $\frac{c^{\prime\,\text{ten}}}{c^{\text{ten}}}$ are presented in Table \ref{tabcaap} (we assume here, for simplicity, that $K_{1A}\approx K_{1}(1400)$). 

Thus, in this case the decay of the tensor mesons does not fix the strength for the axial-tensor since two equivalent terms are present in the original Lagrangian  of Eq. (\ref{chiralag-vp}), and their sum/difference appears for tensors and axial-tensors, respectively.
Nevertheless, it is reasonable to expect that $c^{\prime\,\text{ten}}$ is of the same order of $c^{\text{ten}}$. In general, rather small decay widths are therefore expected. 
Within this context, it is worth to mention that our prediction for the $\rho_2\rightarrow a_1(1260)\pi$ is quite similar to the prediction in Ref. \cite{Guo:2019wpx} for the case of $\frac{c^{\prime\,\text{ten}}}{c^{\text{ten}}}\simeq 1$ .	 

\begin{table}[h] 
		\centering
		\renewcommand{\arraystretch}{1.1}
		\begin{tabular}[c]{|l|c|c|}
			\hline
			Decay process (in model) & eLSM (MeV) & $\kappa_{a_2ap\,,i}$ \\
			\hline \hline
			$\,\;\rho_2(?) \longrightarrow a_1(1260)\, \pi$ & $ \simeq 13 \Big(\frac{c^{\prime\,\text{ten}}}{c^{\text{ten}}}\Big)^2 $ & $2\Big(\frac{\phi_N}{4}\Big)^2$ \\
			\hline
			$\,\;K_{2,A} \longrightarrow a_1(1260)\, K$ & $\simeq  0.1\Big(\frac{c^{\prime\,\text{ten}}}{c^{\text{ten}}}\Big)^2$ & $3\Big(\frac{\sqrt{2}\phi_S}{8}\Big)^2$ \\
			\hline
			$\,\;K_{2,A} \longrightarrow \bar{K}_{1,A}\, \pi $ & $\simeq 11\Big(\frac{c^{\prime\,\text{ten}}}{c^{\text{ten}}}\Big)^2$ & $3\Big(\frac{\phi_N}{8}\Big)^2$ \\
			\hline
			$\,\;K_{2,A} \longrightarrow f_1(1285)\, \bar{K}$ & $\simeq 0.2\Big(\frac{c^{\prime\,\text{ten}}}{c^{\text{ten}}}\Big)^2$ &  $\Big( \frac{\sqrt{2}}{8} (\phi_N\sin{\beta_{A_1}}-\phi_S\cos{\beta_{A_1}})\Big)^2$ \\
			\hline
			$\,\;\omega_{2,S} \longrightarrow \bar{K}_{1,A}\, K + \mathrm{c}.\mathrm{c}.$ & $\simeq 6\Big(\frac{c^{\prime\,\text{ten}}}{c^{\text{ten}}}\Big)^2$ & $4\Big(\frac{\phi_S}{4}\Big)^2$ \\
			\hline
		\end{tabular}
				\caption{Predictions and coefficients for the decay channel $A_{2}\longrightarrow  A_1+P$.}\label{tabcaap}
	\end{table}

\subsubsection{$A_2\longrightarrow T+P$ decay mode}

The last term of the Lagrangian (\ref{eq:mLag}) enables us to make some predictions for the decay of the axial-tensor mesons into a tensor and pseudoscalar pair. Namely, that Lagrangian reduces to the following interaction
\begin{align}
  \mathcal{L}_{a_2tp}= 2ih_3^{\text{ten}} \text{Tr}\Big[A_2^{\mu\nu} (PT_{\mu\nu}\Phi_0 -\Phi_0 T_{\mu\nu}P)\Big]\,,
\end{align}
which leads to the corresponding decay 
\begin{align}
\Gamma_{A_2\longrightarrow T+P}^{tl}(m_{a_2},m_{t},m_{p})=\frac{ (h_3^{\text{ten}})^2|\vec{k}_{t,p}%
|}{2\,m_{a_2}^2\pi}\Big(1+\frac{4|\vec{k}_{t,p}%
|^4}{45m_{t}^4}+\frac{2|\vec{k}_{t,p}%
|^2}{3m_{t}^2}\Big)
\,\kappa_{a_2tp\,,i}\,
\Theta
(m_{a_2}-m_{t}-m_{p})\,.
\end{align}

\begin{table}
	\centering
		\renewcommand{\arraystretch}{1.1}
		\begin{tabular}[c]{|l|c|c|}
			\hline
			Decay process (in model) & eLSM (MeV) & $\kappa_{a_2tp\,,i}$ \\
			\hline \hline
			$\,\;\rho_2(?) \longrightarrow a_2(1320)\, \pi$ & $\approx 88$ & $2\Big(\frac{\phi_N}{4}\Big)^2$  \\
			\hline
			$\,\;K_{2,A} \longrightarrow K_2^{\star}(1430)\,\pi$ & $\approx 49$ & $3\Big(\frac{\sqrt{2}\phi_S}{8}\Big)^2$ \\
				\hline
			$\,\;K_{2,A} \longrightarrow a_2(1320)\,K$ & $\approx 84$ & $3\Big(\frac{\sqrt{2}\phi_N}{8}\Big)^2$ \\
			\hline
			$\,\;K_{2,A} \longrightarrow f_2(1270) \, K $ & $\approx 4$ & $\Big(\frac{\phi_N\cos{\beta_T}-2\phi_S\sin{\beta_T}}{8}\Big)^2$\\
	\hline
			$\,\;\omega_{2,S} \longrightarrow K_2^{\star}(1430)  \, K  + \mathrm{c}.\mathrm{c}.$ & $\approx 15$ & $4\Big(\frac{\phi_S}{8}\Big)^2$ \\
					\hline
		\end{tabular}
				\caption{Predictions and coefficients for the decay channel $A_2\rightarrow T+P$.}\label{a2tp}
	\end{table}
	
The theoretical results are reported in Table \ref{a2tp}. As it is visible, quite sizable decays are obtained in some channels, which are then promising for future investigations of the axial-tensor states.  Note, in \cite{Guo:2019wpx} the decay rate of $\rho_2(?) \longrightarrow a_2(1320)\, \pi$ is calculated to be about $200\,\text{MeV}$.

In the end, it should be stressed that the sum of the various decay rates for the axial-tensor mesons generates quite large decay widths for these states, even by taking the lowest side of the given ranges. Nevertheless, the discovery of the missing axial-tensor states seems possible, especially by looking into the decay channels described in this subsection.

\section{Conclusions}
\label{sec:Conclusion}
In this work we have studied the well known lightest tensor mesons and their (still poorly known) chiral partners, the axial-tensor mesons, in the framework of a chiral model for low-energy QCD, the eLSM.

The masses as well as the strong and radiative decays of the ground-state tensor nonet fit well into the quark-antiquark picture, thus confirming the assignment of the quark model review of the PDG. 
Thanks to chiral symmetry, the parameters determined in the tensor sector allow to make predictions for masses and decays for the basically unknown ground-state axial-tensor resonances (the states $\rho_2$, $\omega_{2,N} \simeq \omega_{2}$, and $\omega_{2,S} \simeq \phi_2$ were not yet seen in experiments; only the kaonic state $K_2(1820)$ is listed in the quark model review of the PDG). 
Even though the uncertainties are large -the widths are determined solely by the chiral link of tensors to axial-tensors and small changes in some key parameters generate quite sizable variations- the emerging picture is quite clear: the axial-tensor mesons are broad, the largest decay channel is the one into vector-pseudoscalar, followed by tensor-pseudoscalar and axial-vector-pseudoscalar pairs, respectively.  Large decay widths into vector-pseudoscalar pairs are also obtained by LQCD and the corresponding decay ratios from lattice fit well with the ones determined by our model calculations. 

We may conclude that the states $\rho_2$, $\omega_2$, and $\phi_2$ can be discovered in ongoing and future experiments (see e.g. Refs.  \cite{Mezzadri:2015lrw,Marcello:2016gcn,Rizzo:2016idq,Ryabchikov:2019rgx,Ghoul:2015ifw,Zihlmann:2010zz,Proceedings:2014joa,Lutz:2009ff,CLAS:2020ngl,Mathieu:2020zpm,Nys:2018vck,Bibrzycki:2013pja} and refs. therein), even if that shall not be an easy task, due to the expected large decay widths.

The study of partial wave analysis \cite{Shastry:2021asu,Shastry:2021fsk} of the decays of (axial-)tensor mesons represents an interesting direction to further constrain models in the future. Moreover, a detailed investigation of the tensor glueball (whose LQCD predicted mass is about 2.2 GeV \cite{Chen:2005mg}) in a chiral framework is a promising outlook, since the topic of glueballs is a relevant one in low-energy QCD and various tensor mesons in the energy region at about 2 GeV exist.

\section*{Acknowledgement}
We are thankful to Adrian K{\"o}nigstein for fruitful discussions on the tensor mesons. 
S. J.  acknowledges financial support through the project ``Development Accelerator of the Jan Kochanowski University of Kielce'', co-financed by the European Union under the European Social Fund, with no. POWR.03.05. 00-00-Z212 / 18.  A. V. and F. G. acknowledge support from the Polish National Science Centre (NCN) through the OPUS project no
2019/33/B/ST2/00613. F. G. and S. J. acknowledge also support through the NCN OPUS no. 2018/29/B/ST2/02576.


\begin{thebibliography}{10}

\bibitem{Zyla:2020zbs}
{\bfseries Particle Data Group} Collaboration, P.~A. Zyla {\em et~al.},
  ``{Review of particle physics},''
  \href{http://dx.doi.org/10.1093/ptep/ptaa104}{{\em PTEP} {\bfseries 2020}
  no.~8, (8, 2020) 083C01}.

\bibitem{isgur1985}
S.~Godfrey and N.~Isgur, ``Mesons in a relativized quark model with
  chromodynamics,'' \href{http://dx.doi.org/10.1103/PhysRevD.32.189}{{\em Phys.
  Rev. D} {\bfseries 32} (Jul, 1985) 189--231}.

\bibitem{Fischer:2014xha}
C.~S. Fischer, S.~Kubrak, and R.~Williams, ``{Mass spectra and Regge
  trajectories of light mesons in the Bethe-Salpeter approach},''
  \href{http://dx.doi.org/10.1140/epja/i2014-14126-6}{{\em Eur. Phys. J. A}
  {\bfseries 50} (2014) 126}, \href{http://arxiv.org/abs/1406.4370}{{\ttfamily
  arXiv:1406.4370 [hep-ph]}}.

\bibitem{Pelaez:2015qba}
J.~R. Pelaez, ``{From controversy to precision on the sigma meson: a review on
  the status of the non-ordinary $f_0(500)$ resonance},''
  \href{http://dx.doi.org/10.1016/j.physrep.2016.09.001}{{\em Phys. Rept.}
  {\bfseries 658} (2016) 1}, \href{http://arxiv.org/abs/1510.00653}{{\ttfamily
  arXiv:1510.00653 [hep-ph]}}.

\bibitem{Sarantsev:2021ein}
A.~V. Sarantsev, I.~Denisenko, U.~Thoma, and E.~Klempt, ``{Scalar isoscalar
  mesons and the scalar glueball from radiative $J/\psi$ decays},''
  \href{http://dx.doi.org/10.1016/j.physletb.2021.136227}{{\em Phys. Lett. B}
  {\bfseries 816} (2021) 136227},
  \href{http://arxiv.org/abs/2103.09680}{{\ttfamily arXiv:2103.09680
  [hep-ph]}}.

\bibitem{Rodas:2021tyb}
{\bfseries Joint Physics Analysis Center} Collaboration, A.~Rodas, A.~Pilloni,
  M.~Albaladejo, C.~Fernandez-Ramirez, V.~Mathieu, and A.~P. Szczepaniak,
  ``{Scalar and tensor resonances in $J/\psi $ radiative decays},''
  \href{http://dx.doi.org/10.1140/epjc/s10052-022-10014-8}{{\em Eur. Phys. J.
  C} {\bfseries 82} no.~1, (2022) 80},
  \href{http://arxiv.org/abs/2110.00027}{{\ttfamily arXiv:2110.00027
  [hep-ph]}}.

\bibitem{Binosi:2022ydc}
D.~Binosi, A.~Pilloni, and R.-A. Tripolt, ``{Study for a model-independent pole
  determination of overlapping resonances},''
  \href{http://arxiv.org/abs/2205.02690}{{\ttfamily arXiv:2205.02690
  [hep-ph]}}.

\bibitem{Klempt:2022qjf}
E.~Klempt, A.~V. Sarantsev, I.~Denisenko, and K.~V. Nikonov, ``{Search for the
  tensor glueball},''
  \href{http://dx.doi.org/10.1016/j.physletb.2022.137171}{{\em Phys. Lett. B}
  {\bfseries 830} (2022) 137171},
  \href{http://arxiv.org/abs/2205.07239}{{\ttfamily arXiv:2205.07239
  [hep-ph]}}.

\bibitem{Klempt:2021nuf}
E.~Klempt, ``{Scalar mesons and the fragmented glueball},''
  \href{http://dx.doi.org/10.1016/j.physletb.2021.136512}{{\em Phys. Lett. B}
  {\bfseries 820} (2021) 136512},
  \href{http://arxiv.org/abs/2104.09922}{{\ttfamily arXiv:2104.09922
  [hep-ph]}}.

\bibitem{Klempt:2021wpg}
E.~Klempt and A.~V. Sarantsev, ``{Singlet-octet-glueball mixing of scalar
  mesons},'' \href{http://dx.doi.org/10.1016/j.physletb.2022.136906}{{\em Phys.
  Lett. B} {\bfseries 826} (2022) 136906},
  \href{http://arxiv.org/abs/2112.04348}{{\ttfamily arXiv:2112.04348
  [hep-ph]}}.

\bibitem{Guo:2022xqu}
D.~Guo, W.~Chen, H.-X. Chen, X.~Liu, and S.-L. Zhu, ``{Newly observed a0(1817)
  as the scaling point of constructing the scalar meson spectroscopy},''
  \href{http://dx.doi.org/10.1103/PhysRevD.105.114014}{{\em Phys. Rev. D}
  {\bfseries 105} no.~11, (2022) 114014},
  \href{http://arxiv.org/abs/2204.13092}{{\ttfamily arXiv:2204.13092
  [hep-ph]}}.

\bibitem{Piotrowska:2017rgt}
M.~Piotrowska, C.~Reisinger, and F.~Giacosa, ``{Strong and radiative decays of
  excited vector mesons and predictions for a new $\phi(1930)$ resonance},''
  \href{http://dx.doi.org/10.1103/PhysRevD.96.054033}{{\em Phys. Rev. D}
  {\bfseries 96} no.~5, (9, 2017) 054033},
  \href{http://arxiv.org/abs/1708.02593}{{\ttfamily arXiv:1708.02593
  [hep-ph]}}.

\bibitem{Burakovsky:1997ci}
L.~Burakovsky and T.~J. Goldman, ``{Regarding the enigmas of $P$-wave meson
  spectroscopy},'' \href{http://dx.doi.org/10.1103/PhysRevD.57.2879}{{\em Phys.
  Rev. D} {\bfseries 57} (3, 1998) 2879--2888},
  \href{http://arxiv.org/abs/hep-ph/9703271}{{\ttfamily arXiv:hep-ph/9703271}}.

\bibitem{Giacosa:2005bw}
F.~Giacosa, T.~Gutsche, V.~E. Lyubovitskij, and A.~Faessler, ``{Decays of
  tensor mesons and the tensor glueball in an effective field approach},''
  \href{http://dx.doi.org/10.1103/PhysRevD.72.114021}{{\em Phys. Rev. D}
  {\bfseries 72} (2005) 114021},
  \href{http://arxiv.org/abs/hep-ph/0511171}{{\ttfamily arXiv:hep-ph/0511171}}.

\bibitem{Koenigstein:2016tjw}
A.~Koenigstein and F.~Giacosa, ``{Phenomenology of pseudotensor mesons and the
  pseudotensor glueball},''
  \href{http://dx.doi.org/10.1140/epja/i2016-16356-x}{{\em Eur. Phys. J. A}
  {\bfseries 52} no.~12, (12, 2016) 356},
  \href{http://arxiv.org/abs/1608.08777}{{\ttfamily arXiv:1608.08777
  [hep-ph]}}.

\bibitem{Shastry:2021asu}
V.~Shastry, E.~Trotti, and F.~Giacosa, ``{Constraints imposed by the partial
  wave amplitudes on the decays of $J=1,2$ mesons},''
  \href{http://arxiv.org/abs/2107.13501}{{\ttfamily arXiv:2107.13501
  [hep-ph]}}.

\bibitem{Jafarzade:2021vhh}
S.~Jafarzade, A.~Koenigstein, and F.~Giacosa, ``{Phenomenology of $J^{PC}$ =
  $3^{--}$ tensor mesons},''
  \href{http://dx.doi.org/10.1103/PhysRevD.103.096027}{{\em Phys. Rev. D}
  {\bfseries 103} no.~9, (2021) 096027},
  \href{http://arxiv.org/abs/2101.03195}{{\ttfamily arXiv:2101.03195
  [hep-ph]}}.

\bibitem{Parganlija:2012fy}
D.~Parganlija, P.~Kovacs, G.~Wolf, F.~Giacosa, and D.~H. Rischke, ``{Meson
  vacuum phenomenology in a three-flavor linear sigma model with (axial-)vector
  mesons},'' \href{http://dx.doi.org/10.1103/PhysRevD.87.014011}{{\em Phys.
  Rev. D} {\bfseries 87} no.~1, (2013) 014011},
  \href{http://arxiv.org/abs/1208.0585}{{\ttfamily arXiv:1208.0585 [hep-ph]}}.

\bibitem{Parganlija:2016yxq}
D.~Parganlija and F.~Giacosa, ``{Excited Scalar and Pseudoscalar Mesons in the
  Extended Linear Sigma Model},''
  \href{http://dx.doi.org/10.1140/epjc/s10052-017-4962-y}{{\em Eur. Phys. J. C}
  {\bfseries 77} no.~7, (2017) 450},
  \href{http://arxiv.org/abs/1612.09218}{{\ttfamily arXiv:1612.09218
  [hep-ph]}}.

\bibitem{Janowski:2014ppa}
S.~Janowski, F.~Giacosa, and D.~H. Rischke, ``{Is f0(1710) a glueball?},''
  \href{http://dx.doi.org/10.1103/PhysRevD.90.114005}{{\em Phys. Rev. D}
  {\bfseries 90} no.~11, (2014) 114005},
  \href{http://arxiv.org/abs/1408.4921}{{\ttfamily arXiv:1408.4921 [hep-ph]}}.

\bibitem{Giacosa:2016hrm}
F.~Giacosa, J.~Sammet, and S.~Janowski, ``{Decays of the vector glueball},''
  \href{http://dx.doi.org/10.1103/PhysRevD.95.114004}{{\em Phys. Rev. D}
  {\bfseries 95} no.~11, (2017) 114004},
  \href{http://arxiv.org/abs/1607.03640}{{\ttfamily arXiv:1607.03640
  [hep-ph]}}.

\bibitem{Eshraim:2020ucw}
W.~I. Eshraim, C.~S. Fischer, F.~Giacosa, and D.~Parganlija, ``{Hybrid
  phenomenology in a chiral approach},''
  \href{http://dx.doi.org/10.1140/epjp/s13360-020-00900-z}{{\em Eur. Phys. J.
  Plus} {\bfseries 135} no.~12, (2020) 945},
  \href{http://arxiv.org/abs/2001.06106}{{\ttfamily arXiv:2001.06106
  [hep-ph]}}.

\bibitem{Molina:2008jw}
R.~Molina, D.~Nicmorus, and E.~Oset, ``{The rho rho interaction in the hidden
  gauge formalism and the f(0)(1370) and f(2)(1270) resonances},''
  \href{http://dx.doi.org/10.1103/PhysRevD.78.114018}{{\em Phys. Rev. D}
  {\bfseries 78} (2008) 114018},
  \href{http://arxiv.org/abs/0809.2233}{{\ttfamily arXiv:0809.2233 [hep-ph]}}.

\bibitem{Geng:2010kma}
L.~S. Geng, F.~K. Guo, C.~Hanhart, R.~Molina, E.~Oset, and B.~S. Zou, ``{Study
  of the f(2)(1270), f(2)-prime(1525), f(0)(1370) and f(0)(1710) in the J/psi
  radiative decays},'' \href{http://dx.doi.org/10.1140/epja/i2010-10971-5}{{\em
  Eur. Phys. J. A} {\bfseries 44} (2010) 305--311},
  \href{http://arxiv.org/abs/0910.5192}{{\ttfamily arXiv:0910.5192 [hep-ph]}}.

\bibitem{Roca:2012rx}
L.~Roca and E.~Oset, ``{Scattering of unstable particles in a finite volume:
  the case of $\pi \rho$ scattering and the $a_1(1260)$ resonance},''
  \href{http://dx.doi.org/10.1103/PhysRevD.85.054507}{{\em Phys. Rev. D}
  {\bfseries 85} (2012) 054507},
  \href{http://arxiv.org/abs/1201.0438}{{\ttfamily arXiv:1201.0438 [hep-lat]}}.

\bibitem{Mezzadri:2015lrw}
G.~Mezzadri, ``{Light hadron spectroscopy at BESIII},''
  \href{http://dx.doi.org/10.22323/1.234.0423}{{\em PoS} {\bfseries 234} (3,
  2015) 423}. The European Physical Society Conference on High Energy Physics
  (EPS-HEP2015).

\bibitem{Marcello:2016gcn}
{\bfseries BESIII} Collaboration, S.~Marcello, ``{Hadron Physics from
  BESIII},'' \href{http://dx.doi.org/10.7566/JPSCP.10.010009}{{\em JPS Conf.
  Proc.} {\bfseries 10} (2016) 010009}. Proceedings of the 10th International
  Workshop on the Physics of Excited Nucleons (NSTAR2015).

\bibitem{Rizzo:2016idq}
{\bfseries CLAS} Collaboration, A.~Rizzo, ``{The meson spectroscopy program
  with CLAS12 at Jefferson Laboratory},''
  \href{http://dx.doi.org/10.22323/1.253.0060}{{\em PoS} {\bfseries 253} (2,
  2016) 8}. The 8th International Workshop on Chiral Dynamics.

\bibitem{Ryabchikov:2019rgx}
{\bfseries VES group, COMPASS} Collaboration, D.~Ryabchikov, ``{Meson
  spectroscopy at VES and COMPASS},''
  \href{http://dx.doi.org/10.1051/epjconf/201921203010}{{\em EPJ Web Conf.}
  {\bfseries 212} (6, 2019) 03010}. The 12th International Workshop on $e^+ \,
  e^-$ Collisions from Phi to Psi (PhiPsi 2019).

\bibitem{Ghoul:2015ifw}
{\bfseries GlueX} Collaboration, H.~Al~Ghoul {\em et~al.}, ``{First results
  from the GlueX experiment},'' \href{http://dx.doi.org/10.1063/1.4949369}{{\em
  AIP Conf. Proc.} {\bfseries 1735} no.~1, (5, 2016) 020001},
  \href{http://arxiv.org/abs/1512.03699}{{\ttfamily arXiv:1512.03699
  [nucl-ex]}}.

\bibitem{Zihlmann:2010zz}
{\bfseries GlueX} Collaboration, B.~Zihlmann, ``{GlueX a new facility to search
  for gluonic degrees of freedom in mesons},''
  \href{http://dx.doi.org/10.1063/1.3483306}{{\em AIP Conf. Proc.} {\bfseries
  1257} no.~1, (8, 2010) 116--120}.

\bibitem{Proceedings:2014joa}
{\bfseries GlueX} Collaboration, M.~Shepherd, ``{GlueX at Jefferson Lab: a
  search for exotic states ofmatter in photon-proton collisions},''
  \href{http://dx.doi.org/10.22323/1.212.0004}{{\em PoS} {\bfseries 212} (10,
  2014) 12}. 52nd International Winter Meeting on Nuclear Physics.

\bibitem{Lutz:2009ff}
{\bfseries PANDA} Collaboration, M.~F.~M. Lutz {\em et~al.}, ``{Physics
  Performance Report for PANDA: Strong Interaction Studies with Antiprotons},''
  3, 2009.

\bibitem{CLAS:2020ngl}
{\bfseries CLAS} Collaboration, M.~Carver {\em et~al.}, ``{Photoproduction of
  the $f_2(1270)$ meson using the CLAS detector},''
  \href{http://dx.doi.org/10.1103/PhysRevLett.126.082002}{{\em Phys. Rev.
  Lett.} {\bfseries 126} no.~8, (2021) 082002},
  \href{http://arxiv.org/abs/2010.16006}{{\ttfamily arXiv:2010.16006
  [nucl-ex]}}.

\bibitem{Mathieu:2020zpm}
{\bfseries JPAC} Collaboration, V.~Mathieu, A.~Pilloni, M.~Albaladejo,
  L.~Bibrzycki, A.~Celentano, C.~Fern\'andez-Ram\'\i{}rez, and A.~P.
  Szczepaniak, ``{Exclusive tensor meson photoproduction},''
  \href{http://dx.doi.org/10.1103/PhysRevD.102.014003}{{\em Phys. Rev. D}
  {\bfseries 102} no.~1, (2020) 014003},
  \href{http://arxiv.org/abs/2005.01617}{{\ttfamily arXiv:2005.01617
  [hep-ph]}}.

\bibitem{Nys:2018vck}
{\bfseries JPAC} Collaboration, J.~Nys, A.~N. Hiller~Blin, V.~Mathieu,
  C.~Fern\'andez-Ram\'\i{}rez, A.~Jackura, A.~Pilloni, J.~Ryckebusch, A.~P.
  Szczepaniak, and G.~Fox, ``{Global analysis of charge exchange meson
  production at high energies},''
  \href{http://dx.doi.org/10.1103/PhysRevD.98.034020}{{\em Phys. Rev. D}
  {\bfseries 98} no.~3, (2018) 034020},
  \href{http://arxiv.org/abs/1806.01891}{{\ttfamily arXiv:1806.01891
  [hep-ph]}}.

\bibitem{Bibrzycki:2013pja}
L.~Bibrzycki and R.~Kaminski, ``{Tensor meson photoproduction as a final state
  interaction effect},''
  \href{http://dx.doi.org/10.1103/PhysRevD.87.114010}{{\em Phys. Rev. D}
  {\bfseries 87} no.~11, (2013) 114010},
  \href{http://arxiv.org/abs/1306.4882}{{\ttfamily arXiv:1306.4882 [hep-ph]}}.

\bibitem{Kloe2}
G.~Amelino-Camelia {\em et~al.}, ``{Physics with the KLOE-2 experiment at the
  upgraded DA$\phi$NE},''
  \href{http://dx.doi.org/10.1140/epjc/s10052-010-1351-1}{{\em Eur. Phys. J. C}
  {\bfseries 68} (2010) 619--681},
  \href{http://arxiv.org/abs/1003.3868}{{\ttfamily arXiv:1003.3868 [hep-ex]}}.

\bibitem{Feldmann:1998vh}
T.~Feldmann, P.~Kroll, and B.~Stech, ``{Mixing and decay constants of
  pseudoscalar mesons},''
  \href{http://dx.doi.org/10.1103/PhysRevD.58.114006}{{\em Phys. Rev. D}
  {\bfseries 58} (10, 1998) 114006},
  \href{http://arxiv.org/abs/hep-ph/9802409}{{\ttfamily arXiv:hep-ph/9802409}}.

\bibitem{tHooft:1986ooh}
G.~'t~Hooft, ``{How instantons solve the $U(1)$ problem},''
  \href{http://dx.doi.org/10.1016/0370-1573(86)90117-1}{{\em Phys. Rept.}
  {\bfseries 142} no.~6, (9, 1986) 357--387}.

\bibitem{Giacosa:2017pos}
F.~Giacosa, A.~Koenigstein, and R.~D. Pisarski, ``{How the axial anomaly
  controls flavor mixing among mesons},''
  \href{http://dx.doi.org/10.1103/PhysRevD.97.091901}{{\em Phys. Rev. D}
  {\bfseries 97} no.~9, (2018) 091901},
  \href{http://arxiv.org/abs/1709.07454}{{\ttfamily arXiv:1709.07454
  [hep-ph]}}.

\bibitem{Llanes-Estrada:2021evz}
F.~J. Llanes-Estrada, ``{Glueballs as the Ithaca of meson spectroscopy: From
  simple theory to challenging detection},''
  \href{http://dx.doi.org/10.1140/epjs/s11734-021-00143-8}{{\em Eur. Phys. J.
  ST} {\bfseries 230} no.~6, (2021) 1575--1592},
  \href{http://arxiv.org/abs/2101.05366}{{\ttfamily arXiv:2101.05366
  [hep-ph]}}.

\bibitem{Cheng:2006hu}
H.-Y. Cheng, C.-K. Chua, and K.-F. Liu, ``{Scalar glueball, scalar quarkonia,
  and their mixing},'' \href{http://dx.doi.org/10.1103/PhysRevD.74.094005}{{\em
  Phys. Rev. D} {\bfseries 74} (11, 2006) 094005},
  \href{http://arxiv.org/abs/hep-ph/0607206}{{\ttfamily arXiv:hep-ph/0607206}}.

\bibitem{Giacosa:2005zt}
F.~Giacosa, T.~Gutsche, V.~E. Lyubovitskij, and A.~Faessler, ``{Scalar nonet
  quarkonia and the scalar glueball: Mixing and decays in an effective chiral
  approach},'' \href{http://dx.doi.org/10.1103/PhysRevD.72.094006}{{\em Phys.
  Rev. D} {\bfseries 72} (11, 2005) 094006},
  \href{http://arxiv.org/abs/hep-ph/0509247}{{\ttfamily arXiv:hep-ph/0509247}}.

\bibitem{Amsler:1995td}
C.~Amsler and F.~E. Close, ``{Is $f_0 (1500)$ a scalar glueball?},''
  \href{http://dx.doi.org/10.1103/PhysRevD.53.295}{{\em Phys. Rev. D}
  {\bfseries 53} (1, 1996) 295--311},
  \href{http://arxiv.org/abs/hep-ph/9507326}{{\ttfamily arXiv:hep-ph/9507326}}.

\bibitem{Amsler:1995tu}
C.~Amsler and F.~E. Close, ``{Evidence for a scalar glueball},''
  \href{http://dx.doi.org/10.1016/0370-2693(95)00579-A}{{\em Phys. Lett. B}
  {\bfseries 353} no.~2--3, (6, 1995) 385--390},
  \href{http://arxiv.org/abs/hep-ph/9505219}{{\ttfamily arXiv:hep-ph/9505219}}.

\bibitem{Hatanaka:2008gu}
H.~Hatanaka and K.-C. Yang, ``{$K_1 (1270)$ -- $K_1 (1400)$ mixing angle and
  new-physics effects in $B \rightarrow K_1 l^+ l^-$ decays},''
  \href{http://dx.doi.org/10.1103/PhysRevD.78.074007}{{\em Phys. Rev. D}
  {\bfseries 78} (10, 2008) 074007},
  \href{http://arxiv.org/abs/0808.3731}{{\ttfamily arXiv:0808.3731 [hep-ph]}}.

\bibitem{Divotgey:2013jba}
F.~Divotgey, L.~Olbrich, and F.~Giacosa, ``{Phenomenology of axial-vector and
  pseudovector mesons: decays and mixing in the kaonic sector},''
  \href{http://dx.doi.org/10.1140/epja/i2013-13135-3}{{\em Eur. Phys. J. A}
  {\bfseries 49} (10, 2013) 135},
  \href{http://arxiv.org/abs/1306.1193}{{\ttfamily arXiv:1306.1193 [hep-ph]}}.

\bibitem{Cheng:2017pcq}
H.-Y. Cheng and X.-W. Kang, ``{Branching fractions of semileptonic $D$ and
  $D_s$ decays from the covariant light-front quark model},''
  \href{http://dx.doi.org/10.1140/epjc/s10052-017-5170-5}{{\em Eur. Phys. J. C}
  {\bfseries 77} no.~9, (2017) 587},
  \href{http://arxiv.org/abs/1707.02851}{{\ttfamily arXiv:1707.02851
  [hep-ph]}}. [Erratum: Eur.Phys.J.C 77, 863 (2017)].

\bibitem{Guo:2019wpx}
D.~Guo, C.-Q. Pang, Z.-W. Liu, and X.~Liu, ``{Study of unflavored light mesons
  with $J^{PC}=2^{--}$},''
  \href{http://dx.doi.org/10.1103/PhysRevD.99.056001}{{\em Phys. Rev. D}
  {\bfseries 99} no.~5, (2019) 056001},
  \href{http://arxiv.org/abs/1901.03518}{{\ttfamily arXiv:1901.03518
  [hep-ph]}}.

\bibitem{Abreu:2020wio}
L.~M. Abreu, F.~M. da~Costa~J\'unior, and A.~G. Favero, ``{Revisiting the
  tensor $J^{PC} = 2^{--}$ meson spectrum},''
  \href{http://dx.doi.org/10.1103/PhysRevD.101.116016}{{\em Phys. Rev. D}
  {\bfseries 101} no.~11, (2020) 116016},
  \href{http://arxiv.org/abs/2004.10736}{{\ttfamily arXiv:2004.10736
  [hep-ph]}}.

\bibitem{Turkan:2019vnc}
A.~T\"urkan, H.~Da\u{g}, J.~Y. S\"ung\"u, and E.~V. Veliev, ``{Hot medium
  effects on pseudotensor K$_2$(1820) meson},''
  \href{http://dx.doi.org/10.1051/epjconf/201919903007}{{\em EPJ Web Conf.}
  {\bfseries 199} (2019) 03007}.

\bibitem{Sungu:2020azn}
J.~Y. S\"ung\"u, A.~T\"urkan, E.~Sertbakan, and E.~V. Veliev, ``{Axial-tensor
  Meson family at $T\ne 0$},''
  \href{http://dx.doi.org/10.1140/epjc/s10052-020-08439-0}{{\em Eur. Phys. J.
  C} {\bfseries 80} no.~10, (2020) 943},
  \href{http://arxiv.org/abs/2005.11526}{{\ttfamily arXiv:2005.11526
  [hep-ph]}}.

\bibitem{Migdal:1982jp}
A.~A. Migdal and M.~A. Shifman, ``{Dilaton Effective Lagrangian in
  Gluodynamics},'' \href{http://dx.doi.org/10.1016/0370-2693(82)90089-2}{{\em
  Phys. Lett. B} {\bfseries 114} (1982) 445--449}.

\bibitem{Salomone:1980sp}
A.~Salomone, J.~Schechter, and T.~Tudron, ``{Properties of Scalar Gluonium},''
  \href{http://dx.doi.org/10.1103/PhysRevD.23.1143}{{\em Phys. Rev. D}
  {\bfseries 23} (1981) 1143}.

\bibitem{Ko:1994en}
P.~Ko and S.~Rudaz, ``{Phenomenology of scalar and vector mesons in the linear
  sigma model},'' \href{http://dx.doi.org/10.1103/PhysRevD.50.6877}{{\em Phys.
  Rev. D} {\bfseries 50} (1994) 6877--6894}.

\bibitem{Carter:1995zi}
G.~W. Carter, P.~J. Ellis, and S.~Rudaz, ``{An Effective Lagrangian with broken
  scale and chiral symmetry: 2. Pion phenomenology},''
  \href{http://dx.doi.org/10.1016/0375-9474(96)80007-E}{{\em Nucl. Phys. A}
  {\bfseries 603} (1996) 367--386},
  \href{http://arxiv.org/abs/nucl-th/9512033}{{\ttfamily
  arXiv:nucl-th/9512033}}. [Erratum: Nucl.Phys.A 608, 514--514 (1996)].

\bibitem{Koenigstein:2015asa}
A.~Koenigstein, F.~Giacosa, and D.~H. Rischke, ``{Classical and quantum theory
  of the massive spin-two field},''
  \href{http://dx.doi.org/10.1016/j.aop.2016.01.024}{{\em Annals Phys.}
  {\bfseries 368} (5, 2016) 16--55},
  \href{http://arxiv.org/abs/1508.00110}{{\ttfamily arXiv:1508.00110
  [hep-th]}}.

\bibitem{Kopke:1988cs}
L.~Kopke and N.~Wermes, ``{J/psi Decays},''
  \href{http://dx.doi.org/10.1016/0370-1573(89)90074-4}{{\em Phys. Rept.}
  {\bfseries 174} (1989) 67}.

\bibitem{OConnell:1995nse}
H.~B. O'Connell, B.~C. Pearce, A.~W. Thomas, and A.~G. Williams, ``{$\rho -
  \omega$ mixing, vector meson dominance and the pion form-factor},''
  \href{http://dx.doi.org/10.1016/S0146-6410(97)00044-6}{{\em Prog. Part. Nucl.
  Phys.} {\bfseries 39} (1997) 201--252},
  \href{http://arxiv.org/abs/hep-ph/9501251}{{\ttfamily arXiv:hep-ph/9501251}}.

\bibitem{Giacosa:2021mbz}
F.~Giacosa, A.~Okopi\'nska, and V.~Shastry, ``{A simple alternative to the
  relativistic Breit\textendash{}Wigner distribution},''
  \href{http://dx.doi.org/10.1140/epja/s10050-021-00641-2}{{\em Eur. Phys. J.
  A} {\bfseries 57} no.~12, (2021) 336},
  \href{http://arxiv.org/abs/2106.03749}{{\ttfamily arXiv:2106.03749
  [hep-ph]}}.

\bibitem{Johnson:2020ilc}
{\bfseries Hadron Spectrum} Collaboration, C.~T. Johnson and J.~J. Dudek,
  ``{Excited $J^{--}$ meson resonances at the SU(3) flavor point from lattice
  QCD},'' \href{http://dx.doi.org/10.1103/PhysRevD.103.074502}{{\em Phys. Rev.
  D} {\bfseries 103} no.~7, (2021) 074502},
  \href{http://arxiv.org/abs/2012.00518}{{\ttfamily arXiv:2012.00518
  [hep-lat]}}.

\bibitem{Shastry:2021fsk}
V.~Shastry, ``{Ratios of Partial Wave Amplitudes in the Decays of $J=1$ and
  $J=2$ Mesons},'' in {\em {19th International Conference on Hadron
  Spectroscopy and Structure}}.
\newblock 12, 2021.
\newblock \href{http://arxiv.org/abs/2112.13221}{{\ttfamily arXiv:2112.13221
  [hep-ph]}}.

\bibitem{Chen:2005mg}
Y.~Chen {\em et~al.}, ``{Glueball spectrum and matrix elements on anisotropic
  lattices},'' \href{http://dx.doi.org/10.1103/PhysRevD.73.014516}{{\em Phys.
  Rev. D} {\bfseries 73} (1, 2006) 014516},
  \href{http://arxiv.org/abs/hep-lat/0510074}{{\ttfamily
  arXiv:hep-lat/0510074}}.

\end{thebibliography}

\end{document}